\documentclass[journal]{IEEEtran}

\usepackage{enumerate}
\usepackage{times}

\usepackage[final]{graphicx}
\usepackage[reqno]{amsmath}
\usepackage{amsfonts}
\usepackage{multicol}
\usepackage{times,amsmath,epsfig}
\usepackage{latexsym,amssymb}
\usepackage{cite}
\usepackage{subfigure}
\usepackage{amsmath}
\usepackage{amssymb}
\usepackage{array}
\usepackage[noend,ruled,vlined]{algorithm2e}
\usepackage{color}
\usepackage{psfrag}
\usepackage{epsfig}

\newtheorem{pro}{Proposition}
\newtheorem{lemma}{Lemma}

\newtheorem{theorem}{Theorem}

\newcommand{\mb}{\mathbf}

\newcommand{\ba}{\mathbf{a}}
\newcommand{\bb}{\mathbf{b}}

\newcommand{\bd}{\mathbf{d}}

\newcommand{\bq}{\mathbf{q}}

\newcommand{\bu}{\mathbf{u}}

\newcommand{\bx}{\mathbf{x}}
\newcommand{\by}{\mathbf{y}}
\newcommand{\bz}{\mathbf{z}}

\newcommand{\E}{\mathbb{E}}

\newcommand{\st}{\mbox{s.t.}}

\newcommand{\define}{{\triangleq}}
\newcommand{\argmin}{\arg\!\min}

\newcommand{\opt}{\textrm{opt}}
\newcommand{\ie}{i.e.}
\newcommand{\eg}{e.g.}

\ifx\BlackBox\undefined
\newcommand{\BlackBox}{\rule{1.5ex}{1.5ex}}  
\fi

\ifx\proof\undefined
\newenvironment{proof}{\par\noindent{\em Proof:\ }}{\hfill\BlackBox\\[.0mm]}
\fi

\graphicspath{{./Figures/}}

\allowdisplaybreaks

\begin{document}

\title{Distributed Real-Time  Power Balancing \\in Renewable-Integrated Power Grids\\ with Storage and Flexible Loads}

\author{Sun Sun, \IEEEmembership{Student Member,~IEEE}, Min Dong, \IEEEmembership{Senior Member,~IEEE}, and Ben Liang, \IEEEmembership{Senior Member,~IEEE}\vspace{-0.5cm}
\thanks{Manuscript received December 28, 2014; revised April 8, 2015; accepted
May 25, 2015. This work was supported by the Natural Sciences
and Engineering Research Council (NSERC) of Canada under Discovery
Grant RGPIN/372059-2009 and Discovery Grant RGPIN-2015-05506.
Paper no. TSG-01285-2014.}
\thanks{
Sun Sun and Ben Liang are with the Department of  Electrical and Computer
Engineering, University of Toronto, Toronto, Canada (email: \{ssun,
liang\}@comm.utoronto.ca).}
\thanks{Min Dong is with the Department of Electrical Computer and Software Engineering, University of Ontario Institute of Technology, Toronto, Canada (email:  min.dong@uoit.ca). }}

\markboth{IEEE Transactions on Smart Grid}{}
\maketitle

\begin{abstract}
The large-scale integration of  renewable generation directly affects the reliability of  power grids. We investigate the problem of power balancing in a general renewable-integrated power grid with storage and flexible loads. We consider a power grid that is supplied by one conventional generator (CG) and multiple renewable generators (RGs) each co-located with  storage, and is connected with external markets. An aggregator operates the power grid  to maintain power balance between supply and demand. Aiming at minimizing the long-term system cost, we first propose a real-time centralized power balancing solution, taking into account the uncertainty of the renewable generation, loads, and energy prices. We then provide a distributed implementation algorithm, significantly reducing both computational burden and communication overhead.
We demonstrate that our proposed algorithm is asymptotically optimal as the storage capacity increases and the CG ramping constraint loosens. Moreover,  the  distributed implementation enjoys a  fast convergence rate, and enables each RG and the aggregator to make their own decisions.
Simulation shows that our proposed  algorithm outperforms   alternatives and can achieve near-optimal performance for a wide range of storage capacity.

\end{abstract}


\section{Introduction}\label{sec:intr}
With  increasing environmental  concerns, more and more renewable energy sources such as wind and solar are expected to be integrated into the power grids.
Renewable generation is often intermittent  with limited dispatchability. Thus, its large-scale integration  could upset the balance between supply and  demand, and affect the system reliability\cite{bkeps}.

To mitigate the randomness of renewable generation, one can employ fast-responsive generators such as natural gas, whose services are nevertheless expensive. Alternative solutions  include energy storage and flexible loads, which may be less costly and meanwhile more environmentally friendly\cite{dek10}\cite{ch11}. In particular, storage can be exploited to shift energy across time; many loads, such as thermostatically controlled loads, electric vehicles, and other smart appliances, can be controlled through curtailment or time shift. Together, storage  and  flexible loads enable adaptive energy absorption and buffering to counter the fluctuation in renewable generation.

In this paper, we investigate the problem of power balancing in a general renewable-integrated power grid with storage and flexible loads, through the coordination of  supply, demand, and storage. Practical power systems are typically operated under multiple time scales. To model this, we consider power balancing for each time scale separately (e.g., \cite{sg13}). More precisely, we focus on  energy management within a single time scale and  aim at proposing a distributed  real-time algorithm  for power balancing. Real-time control is mainly  motivated by the unpredictability  of renewable sources,  which can potentially render off-line algorithms inefficient.
The distributed implementation is to reduce the computational burden of the system operator and also to limit the communication requirement.

\begin{table*}[t]
 \renewcommand{\arraystretch}{1}
 \caption{Comparison with existing works}
 \label{tab:lit}
 \centering
 \begin{tabular}{p{2.5cm}|p{.3cm}|p{.3cm}|p{.3cm}|p{.3cm}|p{.3cm}|p{.3cm}|p{.3cm}|p{.3cm}|p{.3cm}|p{.3cm}|p{.3cm}|p{.3cm}|p{.3cm}|p{1cm}}
 \hline
 & \cite{ltcc13}&\cite{ngj12} &\cite{cas13}&\cite{sun14jsp}&\cite{sun14tsg}&\cite{sg13}&\cite{hwr12p}&\cite{xnls13}&\cite{hmn13}&\cite{css13}
 &\cite{gpfk13}&\cite{zgg13}&\cite{sllf13}& Proposed\\
\hline
 Supply management &Y &Y & & & & &Y &Y & & & &Y &Y &Y\\
 \hline
 Demand management & & &Y & & & &Y & &Y &Y &Y &Y &Y &Y\\
 \hline
 Storage management & & & &Y &Y &Y & &Y &Y &Y &Y &Y &Y &Y\\
\hline
 Uncertainty/dynamics &Y &Y &Y &Y &Y &Y &Y &Y &Y &Y &Y &Y &Y &Y\\
 \hline
 Ramping constraint &Y &Y & & & & & &Y & & & &Y & &Y\\
\hline
 Real-time algorithm &Y &Y &Y &Y &Y &Y &Y &Y &Y &Y &Y & &Y &Y \\
\hline
Distributed algorithm & & &Y&Y&&&&&&&Y&Y&&Y\\
\hline
\end{tabular}
\end{table*}

Earlier works on power balancing commonly ignore system uncertainty  by considering a deterministic operational environment.
There are many recent works explicitly incorporating  system uncertainty into  energy management  of  power grids. Due to  page limitation, we are only able to select some representative papers that are more related to our work. These works emphasize on various issues of the system in energy management (see Table \ref{tab:lit} for a summary).
For example, the authors of \cite{ltcc13} and \cite{ngj12} consider supply side management by assuming that all loads are uncontrollable, the authors of \cite{cas13} study demand side management by optimally scheduling  non-interruptible and deferrable loads of individual users, and the authors of \cite{sg13, sun14jsp}, and \cite{sun14tsg}  propose to employ energy storage to clear power imbalance. In some other works, the authors combine supply side and demand side managements \cite{hwr12p}, or supply side and storage managements \cite{xnls13}, or demand side and  storage managements \cite{hmn13,css13,gpfk13}.

Among existing works,
\cite{zgg13} and \cite{sllf13} are mostly related to our work, in which
all three types of energy management (i.e., supply, demand, and storage) are jointly considered  for power balancing. However, in \cite{zgg13}, although the uncertainty of the renewable generation is considered and characterized by a polyhedral set, the uncertainty of the loads and energy prices  is ignored.
Moreover, the algorithm is designed for off-line use such as in  day-ahead scheduling, and therefore cannot be implemented in real time.
In \cite{sllf13}, a real-time algorithm is proposed to minimize the  cost of a conventional generator (CG) only. Furthermore, the ramping constraint of the CG  is not considered in the algorithm design.
As we will see in this paper, the incorporation of such a constraint can significantly complicate the analysis of the real-time algorithm. In addition, the energy management there is performed centrally  by a system operator.

In this paper, we  include all issues listed in Table \ref{tab:lit} when studying the problem of power balancing.  In particular, we consider a general power grid supplied by a CG and multiple RGs, and  each RG is co-located with an energy storage unit. An aggregator operates the grid by coordinating  supply,  demand, and storage units to maintain the power balancing. Our goal is to minimize   the long-term system cost subject to  the  operational constraints and the quality-of-service requirement of  flexible loads.

Our formulated optimization problem is stochastic in nature, and is technically challenging especially for real-time control. First, owing to the practical operational constraints,  such as the finite storage capacity and  the CG ramping constraint, the control actions are coupled over time, which complicates the real-time decision making. Second,   centralized control of  a potentially large number of RGs by the aggregator may lead to large communication overhead and heavy computation. To overcome the first difficulty, we leverage Lyapunov optimization\cite{bkneely} and develop special techniques to tackle our problem. To address the second challenge, we exploit the structure of the optimization problem and employ the  alternating direction method of multipliers (ADMM)\cite{adboyd}   to offer a distributed algorithm.
Our main contribution is summarized as follows.
\begin{itemize}
\item
We formulate a stochastic optimization problem for  power balancing by taking into account all design issues listed  in Table \ref{tab:lit}.
\item We propose a distributed real-time algorithm for the power balancing optimization problem. We
characterize the performance gap of the proposed algorithm  away from an optimal algorithm, and show that the proposed algorithm is  asymptotically optimal as the storage capacity  increases and the CG ramping constraint  loosens. The algorithm can be implemented in a distributed way, by which each RG and the aggregator can make their own decisions. The distributed implementation enjoys a fast convergence rate and requires limited communication between the aggregator and each RG.
\item
We compare the proposed algorithm with  alternative algorithms by simulation. We show that  our proposed algorithm outperforms the alternatives and is near-optimal  even with small energy storage.
\end{itemize}

Energy storage has been used widely in power grids for combating the variability of renewable generation. A large amount of  works have been reported in literature on storage control and the assessment of its role in renewable integration (e.g., \cite{qcyr14b,kmtz11,kp11,prod12,bb14}).   Compared with these references, this paper focuses on the problem of power balancing, and additionally includes the control of flexible loads in  energy management. A traditional approach for storage control is to formulate the problem as a linear-quadratic regulator (LQR) (e.g., \cite{kmtz11}).  Compared with the Lyapunov optimization approach employed in this paper, the LQR approach is different in terms of its application and the derivation of the control action at each time step. Specifically, the LQR approach applies when the system states evolve according to a set of linear equations and the objective function is quadratic. Obtaining the optimal control action analytically is generally hard and requires system statistics. In contrast, the Lyapunov optimization approach  has no such  requirements on the problem structure, and can additionally deal with long-term time-averaged constraints. Furthermore, in the Lyapunov optimization approach, the control action at each time step is derived by solving an optimization problem with no need for system statistics.

A preliminary version of this work has been presented in \cite{sun14sgc}. In this paper, we  significantly extend \cite{sun14sgc} in two ways: first, we offer a distributed algorithm for practical implementation; second, we provide more in-depth performance analysis of the proposed algorithm both theoretically and numerically, and reveal insights into the interactions of   supply, demand and storage units in maintaining the power balancing of  a grid.

The remainder of this paper is organized as follows.  In Section \ref{sec:sys}, we
describe the system model and formulate the  problem of power balancing.  In Section \ref{sec:real}, we propose  a real-time algorithm and analyze its performance theoretically.  In
Section \ref{sec:dis}, we provide a distributed algorithm for solving the real-time problem. In Section \ref{sec:sim}, we present simulation results.
Finally, we conclude and discuss some future directions in Section \ref{sec:con}. The main symbols used in this paper are summarized in Table \ref{tab:sym}.

\begin{table}[t]
	\renewcommand{\arraystretch}{1.3}
	\caption{List of Main Symbols}
	\label{tab:sym}
	\centering
	\begin{tabular}{|p{0.8cm}|p{7.2cm}|}
		\hline
		$N$ & number of  RGs\\
		\hline
		$l_{b,t}$ & requested amount of  base loads during time slot $t$\\
		\hline
		$l_{f,t}$ & requested amount of  flexible loads during time slot $t$\\
		\hline
		$l_{m,t}$ &  total amount of  satisfied loads during time slot $t$\\
		\hline
		$\alpha$ & portion of  unsatisfied flexible loads\\
		\hline
		$a_i$ &  renewable generation amount of the $i$-th RG during time slot $t$\\
		\hline
		$a_{i,\max}$ &  maximum amount of renewable generation  of the $i$-th RG during each time slot\\
		\hline
		$x_{i,t}$ & charging/discharging amount of the $i$-th storage unit during time slot $t$\\
		\hline
		$|x_{i,\min}|$ & maximum discharging amount\\
		\hline
		$x_{i,\max}$ & maximum charging amount\\
		\hline
		$b_{i,t}$ & contributed energy amount by the $i$-th RG during time slot $t$\\
		\hline
		$s_{i,t}$ & energy state of the $i$-th storage unit at the beginning of time slot $t$\\
		\hline
		$D_i(\cdot)$ & degradation cost function of the $i$-th storage unit\\
		\hline
		$g_t$ & output of the CG during time slot $t$\\
		\hline
		$g_{\max}$ & maximum  output of the CG during each time slot\\
		\hline
		$r$ & ramping coefficient\\
		\hline
		$C(\cdot)$ & generation cost function of the CG\\
		\hline
		$p_{b,t}$  & unit buying price of  external energy markets at time slot $t$\\
		\hline
		$p_{s,t}$  & unit selling price of  external energy markets at time slot $t$\\
		\hline
		$e_{b,t}$ & amount of  energy bought from  external energy markets during time slot $t$\\
		\hline
		$e_{s,t}$ & amount of  energy sold to  external energy markets during time slot $t$\\
		\hline
		$\bq_t$ & system states at time slot $t$\\
		\hline
		$\bu_t$ & control actions at time slot $t$\\
		\hline
		$w_t$ & system cost at time slot $t$\\
		\hline
	\end{tabular}
\end{table}

\section{System Model and Problem Statement}\label{sec:sys}
\subsection{System Model}\label{subsec:sm}
\begin{figure}[h]
\begin{center}
\includegraphics[height=1.5in,width=2.5in]{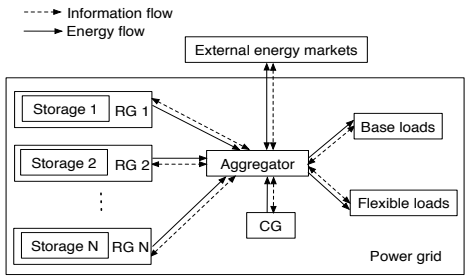}
\end{center}
\vspace{-0.4cm}
\caption{Schematic representation of the considered power grid.}
\label{fig:sys}
\end{figure}

As shown in Fig. \ref{fig:sys},
we consider a power grid supplied by one CG (\eg, nuclear, coal-fired, or gas-fired generator) and $N$ RGs (\eg, wind or solar generators). Each RG is co-located with one on-site energy storage unit.
The grid is connected to  external energy markets and is operated by an aggregator, who is responsible for satisfying the loads by
managing energy from various sources.
The information flow and the energy flow are  also depicted in Fig. \ref{fig:sys}.
Assume that the system operates in discrete time with  time slot $t\in \{0, 1, 2, \cdots\}$.  For notational simplicity, throughout the paper we work with energy units instead of power units.
The details of each component in the power grid are described below.

\subsubsection{Loads}
The loads include  base loads and  flexible loads. The  base loads represent  critical energy demands  such as lighting, which  must be satisfied once requested. The flexible loads here represent  some controllable energy requests that can be partly curtailed if the energy provision cost is high. At time slot $t$, denote the amount of the total requested base loads   by $l_{b,t}\in[l_{b,\min}, l_{b,\max}],$
and the amount of the  total requested flexible loads by $l_{f,t}\in[l_{f,\min},l_{f,\max}]$.
The amounts $l_{b,t}$ and $l_{f,t}$  are generated by users based on their own needs and are considered random.
Let the amount of the total satisfied loads be $l_{m,t}$,  which should satisfy
\begin{align}\label{conl}
		l_{b,t}\le l_{m,t}\le l_{b,t}+l_{f,t}.
\end{align}

The control of  flexible loads needs to meet certain quality-of-service requirement. In this work, we impose an upper bound on the portion of  unsatisfied flexible loads. Formally,
we introduce a long-term time-averaged constraint
\begin{align}\label{conlt}
	\limsup_{T\to\infty}\frac{1}{T}\sum_{t=0}^{T-1}\E\left[\frac{l_{b,t}+l_{f,t}-l_{m,t}}{l_{f,t}}\right] \le \alpha
\end{align}
where $\alpha\in[0,1]$ is a pre-designed threshold with a small value indicating a tight quality-of-service requirement.

\subsubsection{RG and On-Site Storage}
At the $i$-th RG, denote the  amount of the renewable generation
during time slot $t$ by $a_{i,t}\in[0, a_{i,\max}]$, where $a_{i,\max}$ is the maximum generated energy amount.   Due to the stochastic nature of  renewable sources,  $a_{i,t}$ is random.

We assume that each RG is co-located with one on-site energy storage unit capable of charging and discharging. Denote the charging or discharging energy amount of the $i$-th storage unit during time slot $t$ by $x_{i,t}$, with $x_{i,t}>0$ (resp. $x_{i,t}<0$) indicating charging (resp. discharging). Because of the battery design and hardware constraints, the value of $x_{i,t}$ is bounded as follows:
\begin{align}\label{conx}
	x_{i,\min}\le x_{i,t}\le x_{i,\max}, \quad (x_{i,\min}<0<x_{i,\max})
\end{align}
where $|x_{i,\min}|$ and $x_{i,\max}$  represent the maximum discharging and charging amounts, respectively.
For the $i$-th storage unit, denote its energy state at the beginning of time slot $t$ by $s_{i,t}$.
Due to charging and discharging operations, the evolution of  $s_{i,t}$  is given by\footnote{In this work we use a simplified energy storage model. The mathematical framework  carries over when other modeling factors such as  charging efficiency,  discharging efficiency, and storage efficiency are considered.}
\begin{align}\label{st}
s_{i,t+1}   = s_{i,t} +x_{i,t}.
\end{align}
Furthermore, the battery capacity and  operational constraints require the energy state $s_{i,t}$  be bounded as follows:
\begin{align}\label{cons}
	s_{i,\min}\le s_{i,t} \le s_{i,\max}
\end{align}
where $s_{i,\min}$ is the minimum allowed  energy state, and $s_{i,\max}$ is the maximum allowed energy state and can be interpreted as the storage capacity.
It is known that fast charging or discharging can cause battery degradation, which shortens battery lifetime \cite{ram02}.
To model this  cost on  storage, we use $D_{i}(\cdot)$ to represent the degradation cost function associated with the charging or discharging amount $x_{i,t}$.

During every time slot, the RG  supplies energy to the aggregator. Denote the amount of the contributed energy by the $i$-th RG during time slot $t$ by $b_{i,t}$. Since the energy flows of the RG should be balanced, we have
\begin{align}\label{conb}
	b_{i,t}=a_{i,t}-x_{i,t}, \quad b_{i,t}\ge 0.
\end{align}
In particular, if $x_{i,t}>0$ (charging), the contributed energy $b_{i,t}$  directly comes from the renewable generation; if $x_{i,t}<0$ (discharging), $b_{i,t}$ comes from both the renewable generation and the storage unit.

\subsubsection{CG}
Different from the RGs, the energy output of the CG is controllable. Denote $g_t$ as the energy output of the CG during time slot $t$, satisfying
\begin{align}\label{cong}
	0\le g_t\le g_{\max}
\end{align}
where $g_{\max}$ is the maximum amount of the energy output.
Due to the operational  limitations of the CG, the change of the outputs in two consecutive time slots is bounded. This is typically reflected by a \emph{ramping constraint} on the CG outputs \cite{bkmo}.
Assuming that the ramp-up and  ramp-down constraints are identical, we  express the overall ramping constraint as
\begin{align}\label{congr}
	|g_t-g_{t-1}|\le rg_{\max}
\end{align}
where the coefficient $r\in[0,1]$ indicates the tightness of the ramping requirement. In particular,  for $r = 0$, the CG produces a fixed output over time, while for $r=1$, the ramping requirement becomes non-effective.  Furthermore, we denote
the generation cost function of the CG  by $C(\cdot)$.

 \subsubsection{External Energy Markets}
In addition to  the internal energy resources, the  aggregator can resort to the external energy markets if needed. For example, the aggregator can
buy  energy from the external energy markets in the case of  energy deficit,
or sell energy to the  markets in the case of  energy surplus.
At time slot $t$, denote the  unit  prices of the external energy markets for buying  and selling energy by  $p_{b,t}\in[p_{b,\min},p_{b,\max}]$ and $p_{s,t}\in[p_{s,\min}, p_{s,\max}]$, respectively.  To avoid energy arbitrage, the buying price  is assumed to be strictly greater than the selling price, \ie,  $p_{b,t}>p_{s,t}$. The prices  $p_{b,t}$ and $p_{s,t}$ are typically random  due to  unexpected  market behaviors. Denote
\begin{align}\label{cone}
	e_{b,t}\ge0, \quad e_{s,t}\ge 0
	\end{align}
as the amounts of the energy bought from and sold to the external energy markets during time slot $t$, respectively. The overall system balance requirement is
\begin{align}\label{conmbl}
 \textstyle{g_t+e_{b,t}+\sum_{i=1}^Nb_{i,t} = e_{s,t}+l_{m,t}.}
 \end{align}

\subsection{Problem Statement}\label{subsec:ps}
The aggregator operates the power grid and aims to minimize the long-term time-averaged system cost by jointly managing  supply, demand, and storage units. With an increasing integration of renewable generation and energy storage into power grids,  the business models of electric utilities are evolving. From  the study in  \cite{bkfuuti},  one suggested model of future electric utilities is termed as ``energy services utility.'' Such utilities are expected to provide similar services as those described in Section \ref{subsec:sm}. Precisely,  besides serving loads, these utilities would actively provide a platform for demand response, manage  generation assets, and coordinate energy sales with external energy markets.

We define the control actions at time slot $t$ by \[\bu_t \define \left[\bb_t,  \bx_t, l_{m,t}, g_t, e_{b,t}, e_{s,t}\right]\]
where $\bb_t \define [b_{1,t},\cdots, b_{N,t}]$ and $\bx_t \define [x_{1,t},\cdots, x_{N,t}]$.
The system cost at time slot $t$ includes the costs of  all RGs and the CG, and the cost for exploiting the external energy markets,  given by\footnote{For the RGs and the CG, the payment for supplying energy could be settled by additional contracts offered by the aggregator, or be calculated based on the actual provided energy. For these cases,  the payment is transferred inside the system hence not affecting the system-wide cost.}:
\begin{align*}
w_t \define   C(g_t)+p_{b,t}e_{b,t}-p_{s,t}e_{s,t} +\sum_{i=1}^ND_{i}(x_{i,t}).
\end{align*}
Based on the  system model described in Section \ref{subsec:sm},  we formulate
the  problem of power balancing as a stochastic optimization problem below.
\begin{align}
	\nonumber
	\textbf{P1}: \;   \min_{\{\bu_t\}}  \quad  &\limsup_{T\to\infty}\;\frac{1}{T}\sum_{t=0}^{T-1}\E[w_t]\quad  \textrm{s.t.} \quad \eqref{conl}-\eqref{conmbl}
	\end{align}
where the expectations in the objective and \eqref{conlt}  are taken over the randomness of the system states $\bq_t\define [\ba_t, l_{b,t}, l_{f,t}, p_{b,t},p_{s,t}]$ where  $\mathbf{a}_t\define [a_{1,t},\cdots,a_{N,t}]$, and the possible randomness of the control actions.

To keep mathematical exposition simple, we assume that the cost functions $C(\cdot)$ and $D_i(\cdot)$ are continuously differentiable and convex. This assumption is mild since many practical costs  can be well approximated by such functions. Denote  the derivatives of $C(\cdot)$ and $D_i(\cdot)$ by $C'(\cdot)$ and $D'_i(\cdot)$, respectively. Based on the assumption, we have the derivative $C'(g_{t})\in[C'_{\min}, C'_{\max}], \forall g_{t}\in[0, g_{\max}]$, and $ D_{i}'(x_{i,t})\in[D'_{i,\min}, D'_{i,\max}], \forall x_{i,t}\in[x_{i,\min},x_{i,\max}]$.

\textbf{Remarks:}  Compared to a practical power system, the model considered in Section \ref{subsec:sm} is simplified, in which power losses, network constraints, and some other practical operational constraints are ignored. Despite the simplifications, we will show that the proposed formulation leads to an implementable control algorithm with a provable performance bound on  suboptimality. For future work, we will consider incorporating more practical power system constraints into the problem formulation.

\section{Real-Time Algorithm for  Power Balancing}\label{sec:real}
Solving  \textbf{P1} is challenging, due to the
stochastic nature of the system, as well as constraints \eqref{conlt}, \eqref{cons}, and \eqref{congr}, resulting in coupled control actions over time.
In this section, we propose a real-time algorithm for \textbf{P1} and analyze its performance theoretically.

\subsection{Description of Real-Time Algorithm}
To propose a real-time algorithm, we employ the Lyapunov optimization approach\cite{bkneely}. Lyapunov optimization  can be used to transform some long-term time-averaged constraints such as \eqref{conlt} into queue stability constraints, and to
provide efficient real-time algorithms for complex dynamic systems.
Unfortunately, the time-coupled constraints \eqref{cons} and \eqref{congr} are not  time-averaged constraints, but  are hard constraints required at each time slot. Therefore, the Lyapunov optimization framework cannot be directly applied. To overcome this difficulty, we take a  relaxation step and propose the following relaxed problem:
\begin{align}
\nonumber
\textbf{P2}: \;   \min_{\{\bu_t\}} \quad & \limsup_{T\to\infty}\;\frac{1}{T}\sum_{t=0}^{T-1}\E[w_t]\\
\nonumber
\textrm{s.t.}  \quad &
\eqref{conl}-\eqref{conx}, \eqref{conb}, \eqref{cong}, \eqref{cone}, \eqref{conmbl},\\
\label{conltx}
& \lim_{T\to\infty}\;\frac{1}{T}\sum_{t=0}^{T-1} \E[x_{i,t}] = 0, \quad \forall i.
\end{align}
Compared with  \textbf{P1},  in \textbf{P2} the  energy state constraints  \eqref{st} and \eqref{cons} are replaced with a new time-averaged constraint \eqref{conltx}, and  the ramping constraint \eqref{congr} is removed.
It can be shown that \textbf{P2} is indeed a relaxation of \textbf{P1} (see Appendix \ref{app:relax}).

The above relaxation step is crucial and enables us to work under the standard Lyapunov optimization framework. However, we emphasize that, giving solution to \textbf{P2} is not our purpose. Instead, the significance of proposing \textbf{P2} is to facilitate the design of a real-time algorithm  for \textbf{P1} and the performance analysis. Note that due to this relaxation, the solution to \textbf{P2} may  be infeasible to \textbf{P1}. Motivated by this concern, we next provide a real-time algorithm which can guarantee that all constraints of  \textbf{P1} are satisfied.

To meet constraint  \eqref{conlt}, we introduce a  virtual queue backlog $J_t$ evolving as follows:
\begin{align}\label{qj}
J_{t+1} = \max\{J_t-\alpha,0\} +\frac{l_{b,t}+l_{f,t}-l_{m,t}}{l_{f,t}}.
\end{align}
From \eqref{qj}, the virtual queue $J_t$ accumulates the portion of  unsatisfied flexible loads.
It can be shown that maintaining the stability of $J_t$ is equivalent to satisfying constraint \eqref{conlt} \cite{bkneely}. We initialize $J_t$ as $J_0 = 0$.

At  time slot $t$,  define a vector $\mb{\Theta}_t \define[s_{1,t},\ldots,s_{N,t}, J_t]$, which consists of the energy states of all storage units and the virtual queue backlog $J_t$. Using $\mb{\Theta}_t$, we define a  Lyapunov function
$L(\mb{\Theta}_t) \define   \frac{1}{2}J_t^2+\frac{1}{2}\sum_{i=1}^N(s_{i,t}-\beta_i)^2,$ where $\beta_i$ is a  perturbation parameter designed for ensuring the boundedness of the energy state, i.e., constraint \eqref{cons}. In addition, we define the one-slot conditional Lyapunov drift as $\Delta(\mb{\Theta}_t) \define \E\left[L(\mb{\Theta}_{t+1})-L(\mb{\Theta}_t)|\mb{\Theta}_t\right]$.  Instead of directly minimizing the system cost objective, we consider
 the drift-plus-cost function given by  $\Delta(\mb{\Theta}_t)+V \E[w_t|\mb{\Theta}_t]$. It  is  a weighted sum of $\Delta(\mb{\Theta}_t)$ and the system cost at time slot $t$  with $V$ serving as the  weight.

In our algorithm design, we first consider an upper bound on the drift-plus-cost function (see Appendix \ref{app:up}  for the upper bound), and then formulate a real-time optimization problem to minimize this upper bound at every time slot $t$. As a result, at each time slot $t$, we have the following optimization problem:
\begin{align}
	\nonumber
	\textbf{P3}:\;\min_{\bu_t} \quad &\left[\sum_{i=1}^NVD_i(x_{i,t})+(s_{i,t}-\beta_i)x_{i,t}\right] + VC(g_t) \\
	\nonumber
	&+ Vp_{b,t}e_{b,t}-Vp_{s,t}e_{s,t}-\frac{J_t}{l_{f,t}}l_{m,t}\\
	\nonumber
	\textrm{s.t.}  \quad &
	\eqref{conl}, \eqref{conx}, \eqref{conb}-\eqref{conmbl}.
\end{align}
We will show in Section \ref{subsec:pa} that the design of the real-time problem \textbf{P3} can lead to some analytical performance guarantee.  Moreover,  to ensure the feasibility of $g_t$, we take a natural step and move the ramping  constraint \eqref{congr}  back into  \textbf{P3}.

Since $D_i(\cdot)$ and $C(\cdot)$ are convex, \textbf{P3} is a convex optimization problem and can be efficiently solved by standard convex optimization software packages.   Denote an optimal solution of \textbf{P3} at time slot $t$ by $\bu_t^* \define \left[\bb_t^*, \bx_t^*, l_{m,t}^*, g_t^*, e_{b,t}^*, e_{s,t}^*\right]$. At each time slot, after obtaining $\bu_t^*$, we update $s_{i,t}, \forall i$, and $J_{t}$ based on their evolution equations.

In the following proposition we prove that, despite the relaxation to \textbf{P2}, by appropriately designing the perturbation parameter $\beta_i$ we can ensure the boundedness of the energy states and hence the feasibility of the control actions $\{\bu_t^*\}$ to  \textbf{P1}.
\begin{pro}\label{pro:beta}
	For the $i$-th storage unit, set the perturbation parameter $\beta_i$ as
\begin{align}\label{betai}
\beta_i \define V(p_{b,\max}+D'_{i,\max})-x_{i,\min}+s_{i,\min}
\end{align}
where $V\in(0, V_{\max}]$  with
\begin{align}\label{vmax}
		V_{\max} \define \min_{1\le i\le N}\left\{\frac{s_{i,\max}-s_{i,\min}+x_{i,\min}-x_{i,\max}}{p_{b,\max}-p_{s,\min}+D'_{i,\max}-D'_{i,\min}}\right\}.
\end{align}
Then the control actions $\{\bu_t^*\}$ derived by solving \textbf{P3} at each time $t$ are feasible to \textbf{P1}.
\end{pro}
\begin{IEEEproof}
	See Appendix \ref{app:probeta}.
\end{IEEEproof}

\textbf{Remarks:} For $V_{\max}$ in \eqref{vmax} to be positive,  the range of the energy state should be  larger than the sum of the maximum charging and discharging amounts.  This is generally true if the length of each time interval is not too long, for example,  up to several minutes.

We summarize the proposed real-time algorithm in Algorithm \ref{alg:rt}. We can see that,  Algorithm \ref{alg:rt} is simple and does not require any  statistics of the system states.  The latter feature is especially desirable in practice, where accurate  statistics  of the system states are difficult to obtain but instantaneous observations are readily available.

\begin{algorithm}[t]
\caption{Real-time algorithm for power balancing.}
\label{alg:rt}
Initialize $J_0 =0$. At each time slot $t$, the aggregator executes the following steps sequentially.
\begin{enumerate}
	\item Observe the system states $\bq_t$, energy states $s_{i,t}, \forall i,$ and queue backlog $J_{t}$.
	\item Solve \textbf{P3} and obtain an optimal solution $\bu_t^*$.
	\item Use $\bu_t^*$ to update $s_{i,t}, \forall i,$ and $J_{t}$ based on \eqref{st} and \eqref{qj}, respectively.
\end{enumerate}
\end{algorithm}

\subsection{Performance Analysis}\label{subsec:pa}
We now analyze the solution provided by Algorithm \ref{alg:rt} with respect to $\textbf{P1}$.
Under Algorithm \ref{alg:rt},
to emphasize the dependency of the cost objective value on the ramping coefficient $r$ and the control parameter $V$, we denote the achieved  cost objective value  by $w^*(r,V)$.   Denote the minimum  cost objective value of \textbf{P1} by $w^{\opt}(r)$, which only depends on $r$.
The main results are summarized in the following theorem.

\begin{theorem}\label{the:iid}
Assume that  the random system states $\bq_t$ of the  grid   are i.i.d. over time. Then under Algorithm \ref{alg:rt} we have
	\begin{enumerate}
		\item
		$\textstyle{w^*(r,V)-w^{\opt}(r)\le (1-r)g_{\max}\max\{p_{b,\max},C'_{\max}\}}$ $+ B/V
		$,
		where $B$ is a constant defined by  $B\define\frac{1}{2}(1+\alpha^2)+ \frac{1}{2}\sum_{i=1}^N\max\{x_{i,\min}^2,x_{i,\max}^2\}$; and
		\item $w^{\opt}(r) \ge w^*(1,V) -B/V$.
	\end{enumerate}
\end{theorem}
\begin{IEEEproof}
See Appendix \ref{app:theiid}.
\end{IEEEproof}

\textbf{Remarks:}
\begin{itemize}
\item
Theorem \ref{the:iid}.1  characterizes an upper bound on the performance gap  away from $w^{\opt}(r)$.  The upper bound has two terms reflecting the ramping constraint and storage capacity limitation.
It indicates that Algorithm \ref{alg:rt} provides an asymptotically optimal solution as the ramping constraint becomes loose (i.e., $r\to 1$) and the control parameter $V$ increases (or the storage capacity $s_{i,\max}$ increases based on the $V_{\max}$ expression in \eqref{vmax}). This is consistent with our intuition. Using this insight, in order to minimize the gap to the minimum  system cost, we should set $V = V_{\max}$ in Algorithm \ref{alg:rt}.
\item
Theorem \ref{the:iid}.2 provides a lower bound on $w^{\opt}(r)$ in terms of the special case where the ramping constraint is loose, i.e., $r=1$.
Since solving \textbf{P1} to obtain the minimum objective value $w^{\opt}(r)$  is difficult, we will use this lower bound as a benchmark  for performance comparison  in simulation. The gap between the performance under Algorithm \ref{alg:rt} and this lower bound serves as an upper bound on the performance gap between  Algorithm \ref{alg:rt} and an optimal control algorithm.
\item The i.i.d. assumption of the system states $\bq_t$ can be relaxed to accommodate   $\bq_t$  evolving based on a finite state irreducible and aperiodic Markov chain. Similar conclusions can be shown, which are omitted  for brevity.
\end{itemize}

In the above analysis, the storage capacity  $s_{i,\max}$ is assumed to be fixed, so that the control parameter $V$ should be upper bounded by $V_{\max}$ in \eqref{vmax} for ensuring the feasibility of the solution.  Alternatively,  if the storage capacity  can be designed, the question is what its value should be in order to achieve  certain required performance. In the following proposition, we provide an answer to this question by giving an upper bound on the energy state $s_{i,t}$ (hence an upper bound on the minimum required energy capacity) for an arbitrary positive $V$ that can be greater than $V_{\max}$.
\begin{pro}\label{pro:st}
For any $V>0$, the energy state $s_{i,t}$ of the $i$-th storage unit at time slot $t$ under
Algorithm \ref{alg:rt}  satisfies $s_{i,t}\in[s_{i,\min}, s_{i,\textrm{up}}]$ where
\begin{align}
\nonumber
s_{i,\textrm{up}}\define &V(p_{b,\max}-p_{s,\min}+D'_{i,\max}-D'_{i,\min})\\
\label{simax}
&+x_{i,\max}-x_{i,\min}+s_{i,\min}.
\end{align}
\end{pro}
\begin{IEEEproof}
	See Appendix \ref{app:prost}.
\end{IEEEproof}

The expression of  $s_{i,\textrm{up}}$ in \eqref{simax} is  informative and reveals some insights into the dependency of the design of the storage capacity on some system parameters.   First,  $s_{i,\textrm{up}}$ increases linearly with the control parameter $V$. Second, $s_{i,\textrm{up}}$ is larger if the energy prices are more volatile or the  marginal degradation cost increases fast. Third, the minimum $s_{i,\textrm{up}}$ is given by $-x_{i,\min}+x_{i,\max}+s_{i,\min}$  if we have $p_{b,\max} = p_{s,\min}$ and  $D'_{i,\max}=D'_{i,\min}$.

Other properties regarding flexible loads and external transactions  are summarized in the following proposition.
\begin{pro}\label{pro:prop}
	Under  Algorithm \ref{alg:rt} the following results hold.
	\begin{enumerate}
		\item The queue backlog $J_t$ is uniformly bounded from above  as $J_t\le Vp_{b,\max}l_{f,\max}+1.$
		\item The amounts of the external  transactions $e_{b,t}^*$ and $e_{s,t}^*$ satisfy $e_{b,t}^*e_{s,t}^* = 0$.
	\end{enumerate}
\end{pro}
\begin{IEEEproof}
See Appendix \ref{app:proprop}.
\end{IEEEproof}

\textbf{Remarks:}
\begin{itemize}
	\item In Proposition \ref{pro:prop}.1, the upper bound of $J_t$ is deterministic and does not change over time.  Moreover, the fact that $J_t$ is upper bounded  implies that the accumulated  portion of  unsatisfied flexible loads is upper bounded.
	\item Proposition \ref{pro:prop}.2 implies that the aggregator does not buy energy from  or sell energy to the external energy markets simultaneously.
\end{itemize}

\subsection{Discussion on Multiple CGs}
In the current system model, apart from multiple renewable generators, we incorporate  one conventional generator (CG) into the supply side. If there are multiple CGs with the same characteristics, i.e., the same maximum output $g_{\max}$, ramping coefficient $r$, and cost function $C(\cdot)$, for mathematical analysis, we can combine them into one generator. In this case,  the current mathematical framework and the performance analysis  apply directly with the combined generator. The output of each individual CG
can then be obtained by  dividing the output of the combined generator equally over all individual ones.  On the other hand, if these CGs have heterogeneous characteristics and  therefore cannot be combined into one, the proposed algorithm can still be used. In particular, in the original problem \textbf{P1}, we would have  constraints \eqref{cong} and \eqref{congr} for each individual generator; the total output of the generators in \eqref{conmbl} is $\sum_{j = 1}^M g_{j,t}$; and the total cost of the generators is $\sum_{j = 1}^MC_j(g_{i,t})$. The resultant relaxed problem \textbf{P2} would be similar to the current one, in which the ramping constraint \eqref{congr} is removed for each individual CG. For the real-time algorithm, the  formulation of the per-slot optimization problem follows the current mathematical framework.  Moreover,  distributed implementation of the algorithm (shown later in Section \ref{sec:dis}) can  be developed using the same approach we propose.

\section{Distributed Implementation of  Real-Time Algorithm}\label{sec:dis}
At each time slot,  our proposed algorithm (Algorithm \ref{alg:rt})  can be implemented by the aggregator centrally. However, the RGs
may not be willing to relinquish direct control of storage or to offer private information to the aggregator. In addition, the computational complexity of centralized control would grow quickly as the number of RGs increases.
In this section, we provide a distributed algorithm for solving \textbf{P3}, by which each RG and the aggregator can make their own control decisions.

\subsection{Distributed Algorithm Design}
To facilitate the algorithm development, we first transform \textbf{P3} into an equivalent problem. For notational simplicity we drop the time index $t$.
We define a new optimization vector $\by \define [y_1, \cdots, y_{N+4}]$, which relates to the optimization variables of \textbf{P3}  by  $y_i = x_i$ for $1\le i\le N, y_{N+1} = l_m, y_{N+2} = -g, y_{N+3} = -e_b,$ and $y_{N+4} = e_s$.   Then, the objective of \textbf{P3} can be rewritten as the sum of certain functions of each $y_i$, which are denoted by $F_i(y_i)$ but whose details are omitted for brevity. In addition, we replace $b_{i}$ in the constrains of  \textbf{P3}  by $a_{i}-y_{i}$ for $1\le i\le N$ based on  constraint \eqref{conb}. Consequently, \textbf{P3} can be rewritten in a generic form
\textbf{P4} below.
\begin{align*}
	\textbf{P4:}\quad &\min_{\by}  \quad \sum_{i=1}^{N+4}F_i(y_i)  \quad \quad \textrm{s.t.} \; y_i \in \mathcal{Y}_i, \forall i,\; \sum_{i=1}^{N+4} y_i= \sum_{i=1}^Na_i
\end{align*}
where the constraint sets $\{\mathcal{Y}_i\}$ are derived from constraints  \eqref{conl}, \eqref{conx}, and  \eqref{conb}-\eqref{cone}, given by $\mathcal{Y}_i \define [x_{i,\min},\min\{a_i,x_{i,\max}\}], i\in\{1,\cdots,N\}, \mathcal{Y}_{N+1} \define [l_b, l_b+l_f], \mathcal{Y}_{N+2} \define \big[-\min\{g_{\max},g_{t-1}+rg_{\max} \}, -\max\{g_{t-1}-rg_{\max},0\}\big], \mathcal{Y}_{N+3} \define (-\infty, 0], $ and $\mathcal{Y}_{N+4} \define [0, +\infty)$.

Next, we introduce an  auxiliary  vector $\bz$ as a copy of $\by$ and further transform \textbf{P4}  into the following equivalent problem.
\begin{align}
	\nonumber
	\textbf{P5:}\;
	\min_{\by, \bz} \quad &\sum_{i=1}^{N+4}\big[F_i(y_i) + \mb{1}(y_i \in \mathcal{Y}_i)\big] +\mb{1}\Big(\sum_{i=1}^{N+4}z_i = \sum_{i=1}^Na_i\Big)\\
	\label{p4eq}
	\st \quad & \by - \bz = 0
\end{align}
where $\mb{1}(\cdot)$ is an indicator function that equals $0$ if the enclosed event  is true and infinity otherwise.  Through the above transformations, the optimization problem \textbf{P5} now fits the two-block form of the alternating direction method of multipliers (ADMM) \cite{adboyd}, enabling us to develop the distributed optimization algorithm.

Following a general  ADMM approach\cite{adboyd}, we  associate the equality constraint  \eqref{p4eq} in \textbf{P5} with  dual variables $\bd \define [d_1, \cdots, d_{N+4}]$.  Denote $y_i^k, z_i^k, $ and $d_i^k$ as the respective variable values at the $k$-th iteration. Then, based on ADMM,  these values  are updated as follows.
\begin{align}
	\label{yiup}
	& y_i^{k+1} = \argmin_{y_i}\left\{ F_i(y_i)+\frac{\rho}{2}\big(y_i-z_i^k+\frac{d_i^k}{\rho}\big)^2|y_i\in\mathcal{Y}_i\right\}, \forall i,\\
	\label{zup}
	& \bz^{k+1} = \argmin_{\bz}\left\{\sum_{i=1}^{N+4}\big(z_i-\frac{d_i^k}{\rho}-y_i^{k+1}\big)^2|\sum_{i=1}^{N+4}z_i= \sum_{i=1}^Na_i\right\},\\
	\label{diup}
	& d_i^{k+1} = d_{i}^k + \rho(y_i^{k+1}-z_i^{k+1}), \forall i
\end{align}
where  $\rho>0$ is a penalty parameter, which needs to be carefully adjusted for good convergence performance\cite{adboyd}.

After further algebraic manipulation (see Appendix \ref{app:deradmm}), we can eliminate the vectors $\bz$ and $\bd$ and simplify the  updates \eqref{yiup}-\eqref{diup}  as follows:
\begin{align}
	\label{yupf}
	& y_i^{k+1} = \argmin_{y_i}\left\{ F_i(y_i)+\frac{\rho}{2}\big(y_i-v_i^k)^2|y_i\in\mathcal{Y}_i\right\}, \forall i,\\
	\label{dup}
	&d^{k+1} = d^k+\rho\left(\overline{y}^{k+1}-\frac{1}{N+4}\sum_{i=1}^Na_i\right).
\end{align}
In \eqref{yupf}, we have
	$v_{i}^k\define y_i^k-\overline{y}^k-\frac{d^k}{\rho}+ \frac{1}{N+4}\sum_{i=1}^Na_i$ where $\overline{y}^k\define\frac{1}{N+4} \sum_{i=1}^{N+4}y_i^k$ and $d^k$ is a scalar updated as in \eqref{dup}.

\textbf{Remarks:} Following the proof of Theorem 2 in \cite{wb13}, we can show   that the above updates lead to a  worst-case convergence rate $O(1/k)$.  Compared with the subgradient-based algorithm, which presents a worst-case convergence rate $O(1/\sqrt{k})$, the proposed distributed algorithm is much faster and thus is well suited for  real-time implementation.

\subsection{Distributed Implementation}
\begin{figure}[t]
	\begin{center}
		\includegraphics[height=1.2in,width=2.7in]{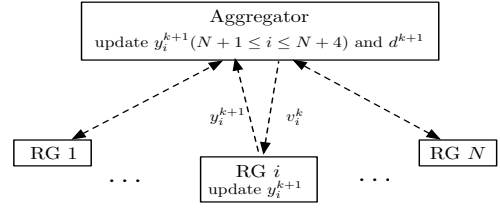}
	\end{center}
	\vspace{-0.4cm}
	\caption{Information flow of distributed implementation.}
	\label{fig:dis}
\end{figure}

Now we discuss the implementation of the proposed distributed algorithm in terms of both computation and communication. In Fig. \ref{fig:dis}, we depict the information flow between the aggregator and the RGs  for the updates in  \eqref{yupf} and \eqref{dup} at the $(k+1)$-th iteration.

Note that the minimization problems in  \eqref{yupf}  can be solved individually at each RG $i$ for $1\le i\le N$,  and at the aggregator  for $N+1\le i\le N+4$, while the update in \eqref{dup} can be computed by the aggregator.
At the initial iteration $k=0$, each RG $i$ needs to send  its renewable generation amount  $a_{i}$ to the aggregator. At each iteration, the aggregator sends a signal $v_{i}^k$ to each RG $i$. Then RG $i$ obtains the update $y_i^{k+1}$ and sends it back to the aggregator.
We  see that, the RGs do not have to release any other  private information to the aggregator, and  the required information exchange  is limited  to one variable in each direction per RG.

Note that the minimization problems in \eqref{yupf} are all strictly convex and admit a unique (and sometimes closed-form) solution. Furthermore, effectively, only one dual variable is  required to be updated in \eqref{dup}. This is because the transformation from \textbf{P3} to \textbf{P4} by introducing  the new optimization vector $\by$ permits all dual variables to share the same updating structure, hence  reducing the number of the effective dual updates  as well as simplifying the calculation.

\section{Simulation Results}\label{sec:sim}
In this section, we evaluate the proposed real-time algorithm and compare it with alternatives  using an idealized but representative power grid setup.

\subsection{Simulation Setup}
Unless otherwise specified, the following parameters are set as default. The length of each time slot is $10$ min.  The amounts of the base loads $l_{b,t}$  and the flexible loads $l_{f,t}$ are uniformly distributed between $5$ and $25$ kWh, and the portion of  unsatisfied flexible loads $\alpha$ is $0.5$. The aggregator is connected with $N=30$ RGs. For each on-site storage unit, we set the maximum discharging and charging amounts to be $1.1$ kWh by assuming that the  discharging and charging  rate is $6.6$ kW (three-phase, level II)\cite{ia09}. Since the model of the degradation cost function of storage is usually  proprietary and unavailable, in simulation, we set $D_i(x) = 10x^2$ as an example. The renewable generation $a_{i,t}$ is uniformly distributed between $0$ and $1.1$ kWh. For the CG, we set the generation cost function to be $C(x) = 8x$, the maximum output $g_{\max} = 50$ kWh, and the ramping coefficient $r =0.1$. The unit buying energy price $p_{b,t}$ is uniformly distributed  between $10$ and $12$ cents/kWh, which is around the current mid-peak energy price in Ontario\cite{onep}. The unit selling energy price $p_{s,t}$ is uniformly distributed between  $4$ and $6$ cents/kWh, which  is slightly below the current off-peak energy price in Ontario\cite{onep}. The control parameter $V$ is set to $1$, $s_{i,\min} = 0$, and $s_{i,\max}$ is given by \eqref{simax}.

\subsection{Benchmark Algorithms}
As discussed in Section \ref{sec:intr},  compared with previous works (e.g., \cite{ltcc13,ngj12,cas13,sun14jsp,sun14tsg,sg13,hwr12p,xnls13,hmn13,css13,gpfk13,zgg13,sllf13}), this paper is built on a more general system model in which all issues listed in Table \ref{tab:lit} are incorporated into the problem formulation. Therefore, mathematically, the problem we study is new and different from all previous ones. As a result, the proposed algorithm cannot be directly compared with the algorithms presented in \cite{ltcc13,ngj12,cas13,sun14jsp,sun14tsg,sg13,hwr12p,xnls13,hmn13,css13,gpfk13,zgg13,sllf13}. To overcome this difficulty, we employ two alternative algorithms  as well as the lower bound on the minimum system cost derived in Theorem \ref{the:iid}.2 for comparison.

The first alternative  is a greedy algorithm, which only minimizes the current system cost. The optimization problem of the greedy algorithm at  time slot $t$ is formulated as follows:
\begin{align*}
 \min_{\bu_t} \quad & w_t\\
\st \quad & \eqref{conx},\eqref{conb}-\eqref{conmbl},\\
& \textstyle{l_{b,t}+(1-\alpha)l_{f,t}\le l_{m,t}\le l_{b,t}+l_{f,t},}\\
& \textstyle{-s_{i,t}\le x_{i,t}\le s_{i,\max}-s_{i,t}.}
\end{align*}

The second alternative is  suggested mainly to show  the  effect of the ramping constraint. In particular, at each  time slot $t$, we solve an optimization problem that is the same as \textbf{P3} except without the ramping constraint \eqref{congr}. Therefore, the resultant CG output may  be  infeasible to \textbf{P1}. To maintain  feasibility,  whenever the CG output  violates the ramping constraint,
 the aggregator only  uses  the external energy markets to augment the CG output. We call it ``naive algorithm'' below.

\subsection{Comparison under  Parameters $V$ and $\alpha$}
\begin{figure}[t]
	\centering
	\includegraphics[height=1.5in,width=2.2in]{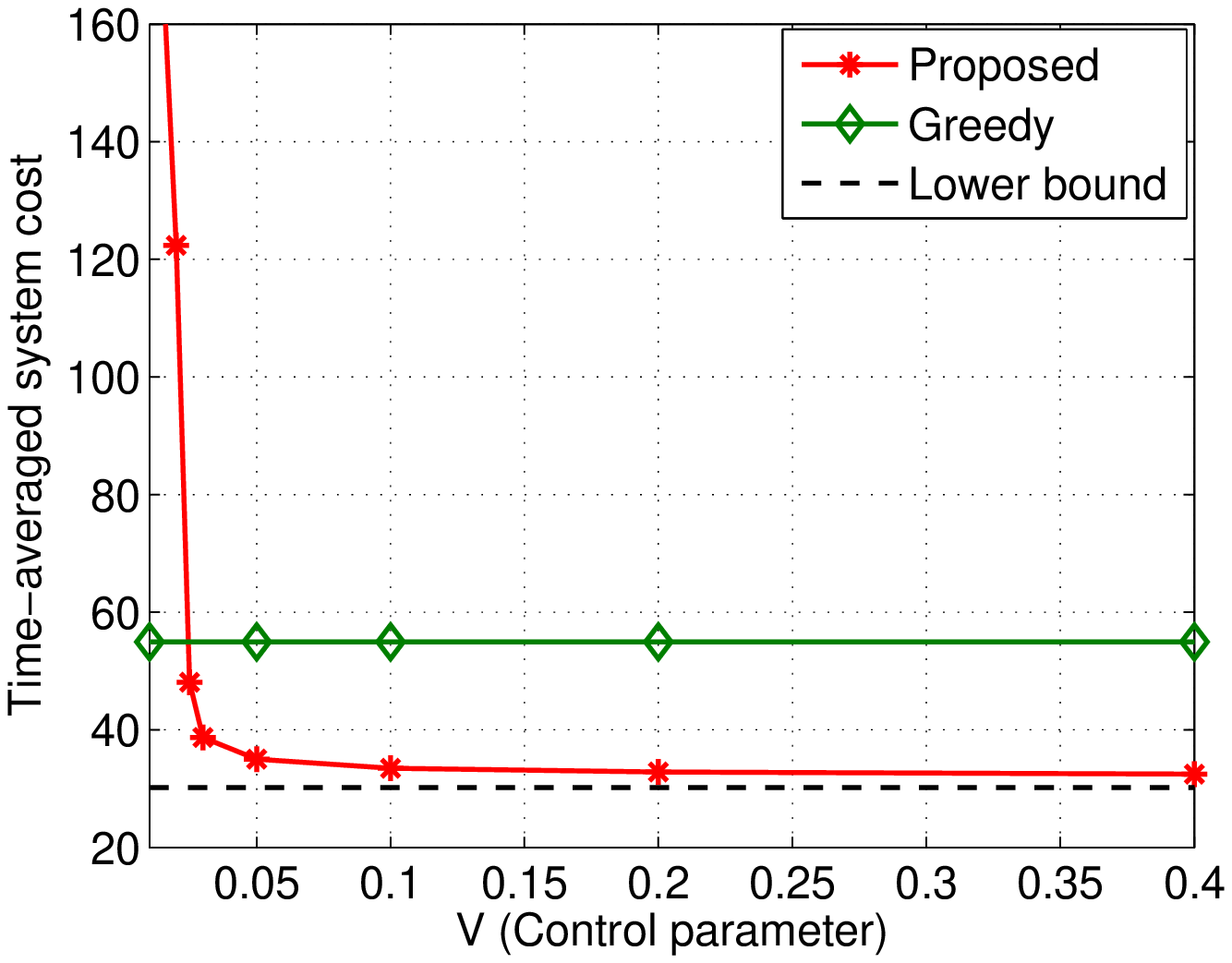}
	\vspace{-0.2cm}
	\caption{System cost vs. control parameter $V$.}
	\label{fig:difvn30}
\centering
\includegraphics[height=1.5in,width=2.2in]{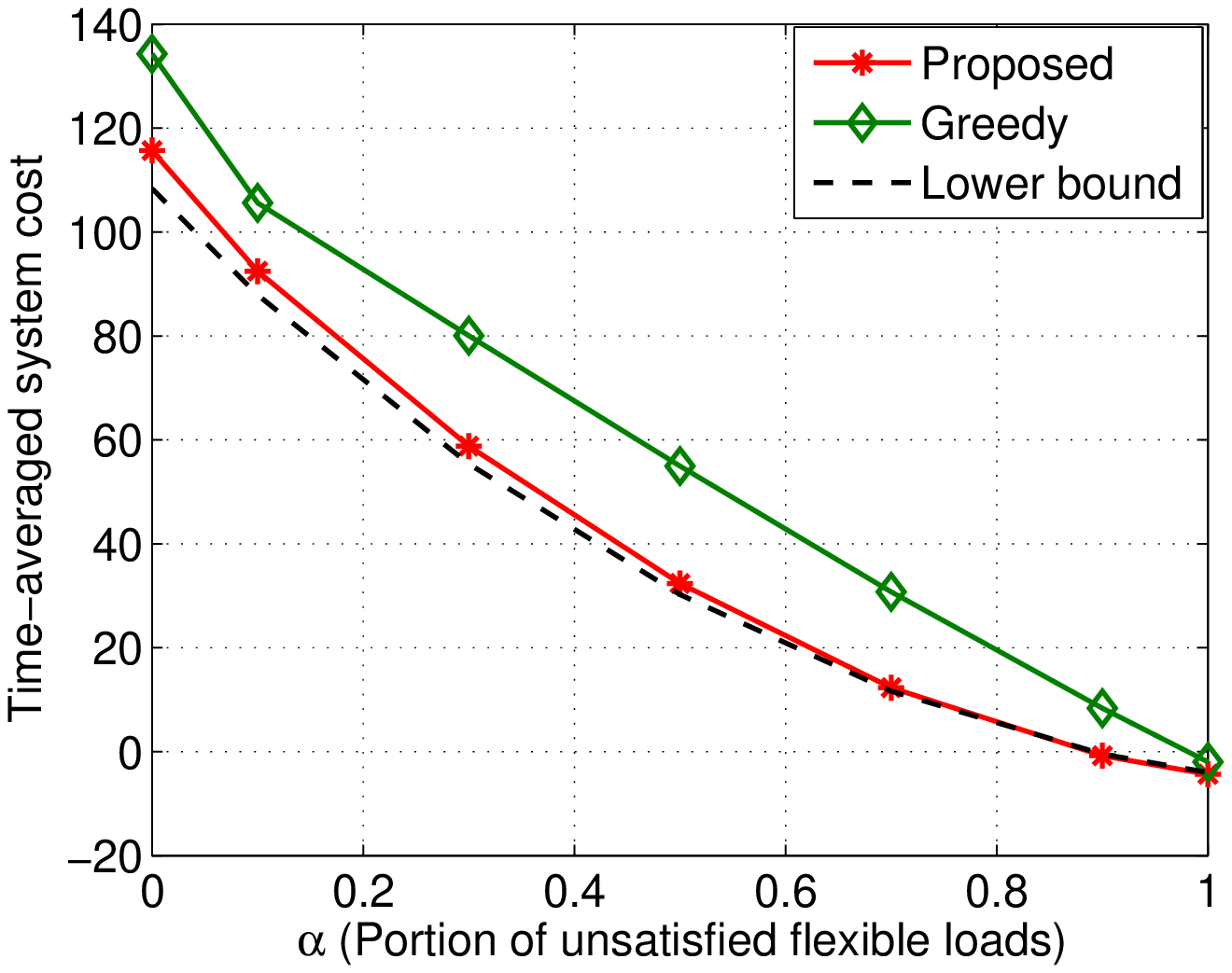}
\vspace{-0.2cm}
\caption{System cost vs. portion of unsatisfied flexible loads $\alpha$.}
\label{fig:difan30}
\centering
\includegraphics[height=1.5in,width=2.2in]{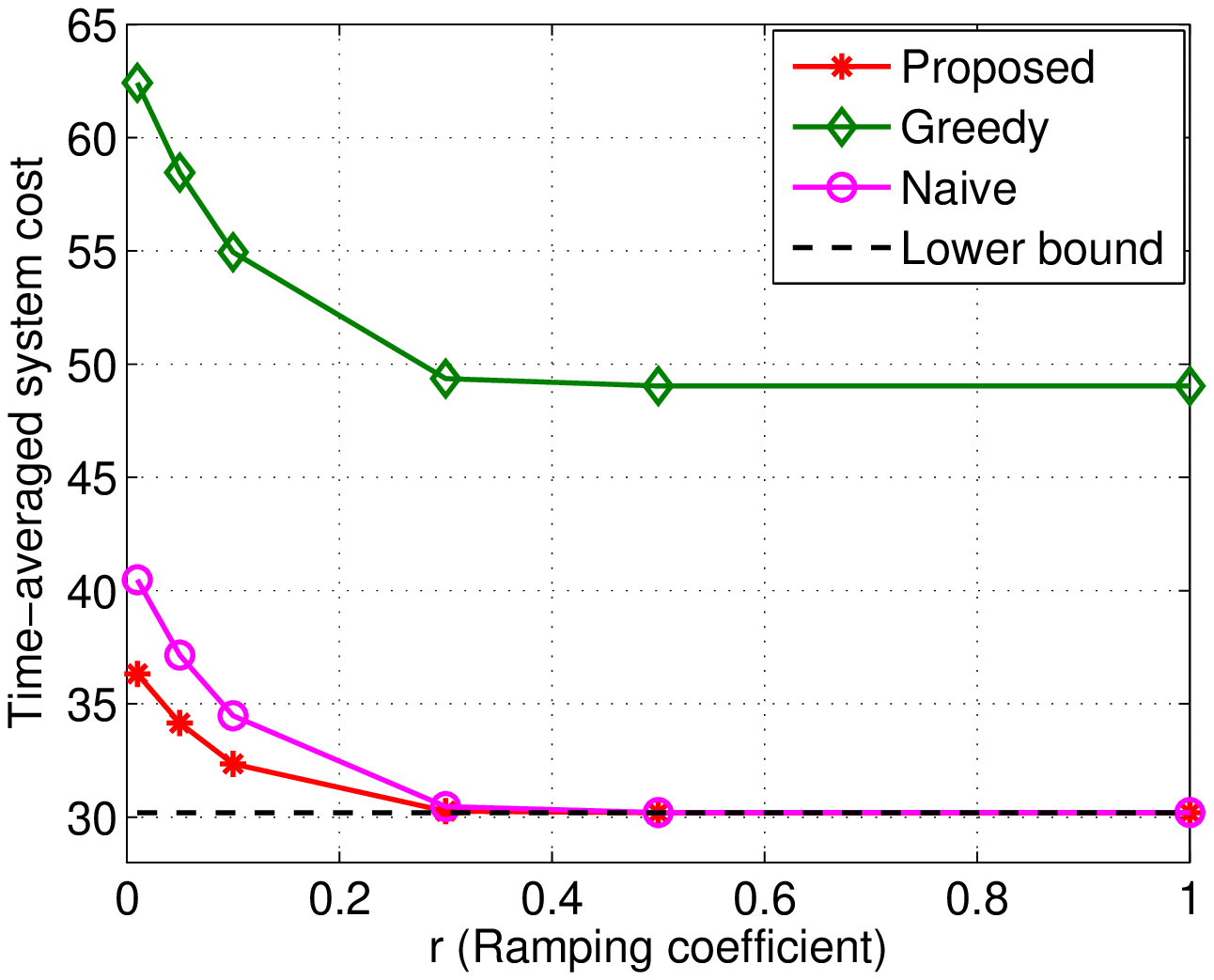}
	\vspace{-0.2cm}
	\caption{System cost vs. ramping coefficient $r$ (small loads).}
	\label{fig:difrn30d5}
\end{figure}

In Fig. \ref{fig:difvn30}, we depict the time-averaged system cost under various values of the control parameter $V$. For the proposed algorithm, the system cost drops quickly and then remains stable as it drops close to the lower bound. This observation demonstrates the efficiency of the algorithm and implies that using small storage  may be enough to achieve near-optimal performance. In contrast,  the performance of the greedy algorithm  barely changes with $V$. In particular, the system cost under the greedy algorithm is about $1.7$ times that under the proposed algorithm  when $V\ge 0.1$.

In Fig. \ref{fig:difan30}, we illustrate the effect of $\alpha$, the portion of  unsatisfied flexible loads. As expected, the system cost goes down as $\alpha$ rises, since less load is  to be satisfied. For the proposed algorithm, the marginal system cost decreases with $\alpha$, which indicates that the benefit of curtailing   loads keeps on falling. We also notice that the greedy algorithm is comparable  with the proposed algorithm for  $\alpha = 1$. But for general cases of $\alpha$,
the proposed algorithm  is observed to have a noticeable advantage. In addition, the proposed algorithm is close to the minimum system cost for all cases.

\subsection{Effect of Ramping Constraint}
In Fig. \ref{fig:difrn30d5} we first consider a scenario with small loads.   The system cost is shown to be non-increasing with respect to the ramping coefficient $r$. This is easy to understand since  a looser ramping constraint implies less usage of the expensive external energy markets.
Furthermore, for all algorithms,  the system cost cannot be decreased any further for $r\ge 0.3$.  This indicates that the CG supply is already sufficient at this point, and therefore a further relaxation of the ramping constraint is unnecessary. We observe that,  the proposed algorithm outperforms both alternatives for all cases. However, the proposed and naive algorithms coincide when $r\ge 0.3$. This  happens because with sufficient supply and a relaxed ramping constraint, the need for augmenting the  CG output in the naive algorithm is small. That is, the control actions under the naive algorithm are consistent with those under the proposed algorithm in most cases.

In Fig. \ref{fig:difrn30d20}, we study a more stressed power grid by
increasing the loads. We assume that $l_{b,t}$ and $l_{f,t}$ are
distributed between $20$ and $40$ kWh.
For the proposed and  naive algorithms, the ramping constraint now has a more noticeable impact. First, the system cost under these two algorithms keeps on dropping for larger $r$, and second, the proposed algorithm always outperforms the naive algorithm. In addition, for  small $r$, the naive algorithm is unsatisfactory as its performance is close to that of the greedy algorithm. This observation  shows the importance of jointly exploiting the system resources, especially under a stressful system environment.

\subsection{Convergence of Distributed  Implementation}

\begin{figure}[t]
	  \centering
	  \includegraphics[height=1.5in,width=2.2in]{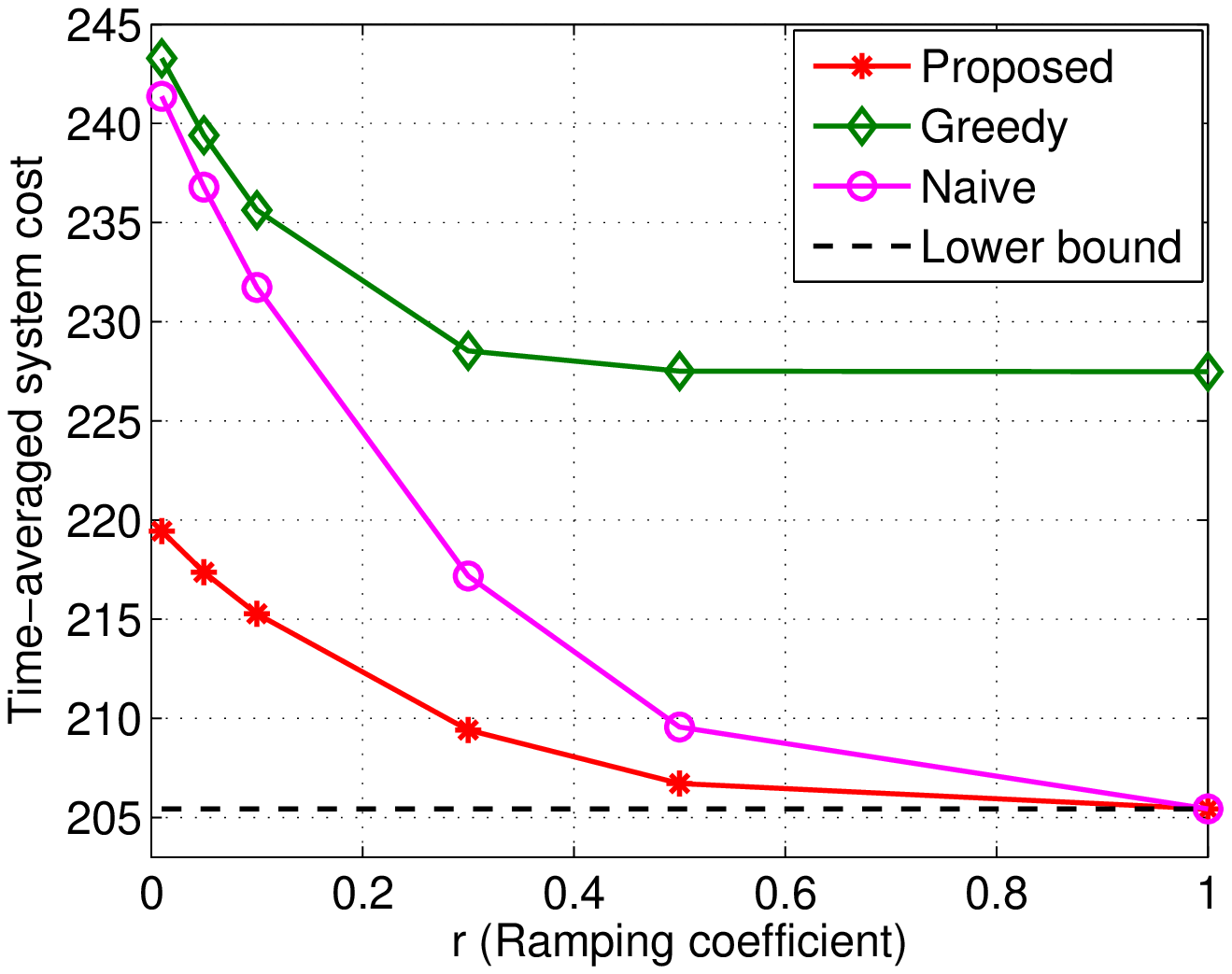}
	  \vspace{-0.2cm}
	  \caption{System cost vs. ramping coefficient $r$  (large loads).}
	  \label{fig:difrn30d20}
	\centering
    \includegraphics[height=1.5in,width=2.2in]{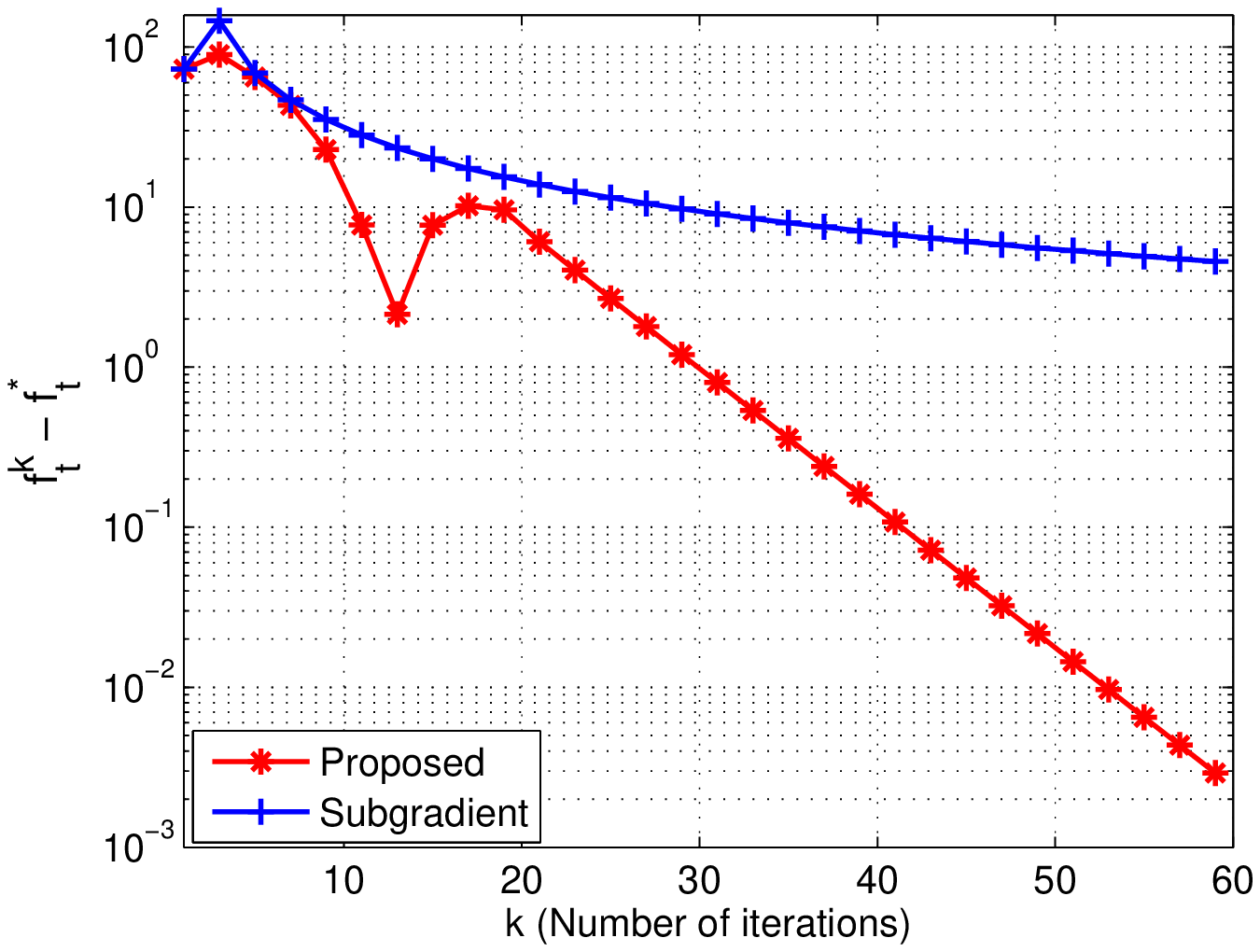}
	\vspace{-0.2cm}
	\caption{Performance gap vs. number of iterations for distributed algorithm.}
	\label{fig:admm}
\end{figure}

In Fig. \ref{fig:admm},  we exhibit the convergence  of the proposed distributed algorithm for a particular system realization. The value of the penalty parameter $\rho$ needs to be adjusted for good convergence performance and is set to $5$ in our case.  For comparison, we also show the convergence of a subgradient algorithm \cite{lps99}. The vertical axis denotes the gap between the value of the objective function and the minimum value of the objective function  of \textbf{P5}.
We  see that, the proposed  algorithm converges fast and exhibits a linear convergence rate, while the subgradient algorithm is slow and exhibits a sublinear convergence rate. Moreover, the fast convergence of the proposed algorithm is observed in general, and we omit the curves of the other system realizations for brevity.

\section{Conclusion and Future Work}\label{sec:con}
We have investigated the problem of power balancing in a renewable-integrated power grid with storage and flexible loads. With the objective of minimizing the system cost, we have proposed a distributed real-time algorithm, which is fast converging and is asymptotically optimal as the storage capacity increases and the ramping constraint of the CG becomes loose.

There are several possible directions for the future work. For example,  first, in the proposed real-time algorithm, only the  current observations of the  system states are employed  in the algorithm design. In reality,  forecasts of the system states (e.g., wind generation, loads, and electricity prices) are usually available within a certain time interval. Therefore, it would be interesting to study how to incorporate these forecasts into the algorithm design and
how these forecasts could improve the  algorithm performance.
Second, the specific implementation of curtailing  the flexible loads is not considered in this paper. How to incentivize  individual customers to participate in such power balancing service and other demand response programs is currently open and  worth further investigation.


\appendices

\section{Proof of Relaxation from \textbf{P1} to \textbf{P2}}\label{app:relax}
Using the  energy state update in \eqref{st}   we can derive  that the left hand side of  constraint \eqref{conltx} equals  the following:
\begin{align}\label{xst}
\lim_{T\to\infty}\;\frac{1}{T}\sum_{t=0}^{T-1} \E[x_{i,t}] = \lim_{T\to\infty}\;\frac{\E[s_{i,T}]}{T}-\lim_{T\to\infty}\;\frac{\E[s_{i,0}]}{T}.
\end{align}
In \eqref{xst}, if $s_{i,t}$ is always bounded, \ie,  constraint \eqref{cons} holds, then the right hand side of  \eqref{xst} equals zero and thus constraint \eqref{conltx} is satisfied. Therefore, \textbf{P2} is  a relaxed problem of \textbf{P1}.

\section{Upper bound on drift-plus-cost function}\label{app:up}
In the following lemma, we show that  the drift-plus-cost function  is upper bounded.
\begin{lemma}\label{lem:dpp}
	For all possible decisions and all possible values of $\mb{\Theta}_t$, in each time slot $t$, the drift-plus-cost function is upper bounded as follows:
	\begin{align}
	\nonumber
	&\Delta(\mb{\Theta}_t) + V \E[w_t|\mb{\Theta}_t]\le B + J_t\E\left[\frac{l_{b,t}+l_{f,t}-l_{m,t}}{l_{f,t}}-\alpha\Big|\mb{\Theta}_t\right]\\
	\label{dppup}
	&\hspace{1cm}+\sum_{i=1}^N(s_{i,t}-\beta_i)\E\big[x_{i,t}|\mb{\Theta}_t\big]+V \E[w_t|\mb{\Theta}_t]
	\end{align}
	where $B$ is a constant and is given by
	$B\define\frac{1}{2}(1+\alpha^2)+ \frac{1}{2}\sum_{i=1}^N\max\{x_{i,\min}^2,x_{i,\max}^2\}.$
\end{lemma}
\begin{IEEEproof}
	Based on the definition of $L(\mb{\Theta}_t)$, the difference
	\begin{align}\label{difl}
	\nonumber
	& L(\mb{\Theta}_{t+1}) - L(\mb{\Theta}_{t})\\
	& = \frac{1}{2}\left[\sum_{i=1}^N(s_{i,t+1}-\beta_i)^2-(s_{i,t}-\beta_i)^2 \right] + \frac{1}{2} (J_{t+1}^2-J_t^2).
	\end{align}
	From the iteration of $J_t$ in \eqref{qj},  $(J_{t+1}^2-J_t^2)$ in \eqref{difl} can be upper bounded as
	\begin{align}\label{difj}
	J_{t+1}^2-J_t^2\le 2J_t\left(\frac{l_{b,t}+l_{f,t}-l_{m,t}}{l_{f,t}}-\alpha\right) + 1+\alpha^2.
	\end{align}
	From the iteration of $s_{i,t}$  in \eqref{st},  $[(s_{i,t+1}-\beta_i)^2-(s_{i,t}-\beta_i)^2]$ in \eqref{difl} can be upper bounded as
	\begin{align}
	\nonumber
	&(s_{i,t+1}-\beta_i)^2-(s_{i,t}-\beta_i)^2\\
	\label{difs}
	&\hspace{-0.5cm}\le 2x_{i,t}(s_{i,t}-\beta_i) + \max\{x_{i,\min}^2,x_{i,\max}^2\}.
	\end{align}
	
	Applying  inequalities  \eqref{difj} and \eqref{difs} to \eqref{difl}, taking the conditional expectation given $\mb{\Theta}_t$, and adding the term $V \E[w_t|\mb{\Theta}_t]$ yields the upper bound in \eqref{dppup}.
\end{IEEEproof}

\section{Proof of Proposition \ref{pro:beta}}\label{app:probeta}
To prove the feasibility under Algorithm \ref{alg:rt}, we are left to show that the long-term constraint \eqref{conlt} and the energy state constraint \eqref{cons} are satisfied.

For constraint \eqref{conlt}, under  the Lyapunov optimization framework, it suffices to show that the virtual queue $J_t$ is mean rate stable, \ie, $ \lim_{T\to\infty}\frac{\E[J_{i,T}]}{T}=0$ (see Section 4.4 in \cite{bkneely}).
Using  Proposition \ref{pro:prop}.1  that $J_t$ is  upper bounded we can easily prove this identity.

To prove that constraint \eqref{cons} is satisfied, we first show the following lemma which gives a sufficient condition for charging or discharging.

\begin{lemma}\label{lem:sit}
	\quad
	\begin{enumerate}
		\item If $s_{i,t}< -x_{i,\min}+s_{i,\min}$, then $x_{i,t}^* = \min\{a_{i,t},x_{i,\max}\}$.
		\item If $s_{i,t}>\beta_i-V(p_{s,\min}+D'_{i,\min})$, then $x_{i,t}^* = x_{i,\min}$.
	\end{enumerate}
\end{lemma}
\begin{IEEEproof}
	To show  Lemma \ref{lem:sit}.1), we first transform \textbf{P3} to an equivalent problem \textbf{P3a)} by eliminating the variables $e_{b,t}$ and $b_{i,t}, \forall i$, and the constant terms.
	\begin{align}
	\nonumber
	\textbf{P3a)}: \min \; &\textstyle{\big[\sum_{i=1}^NVD_i(x_{i,t})+(s_{i,t}-\beta_i)x_{i,t}\big] + VC(g_t)} \\
	\nonumber
	&\hspace{-2cm}\textstyle{+ Vp_{b,t}\big(e_{s,t}+l_{m,t}-g_t+\sum_{i=1}^Nx_{i,t}\big)-Vp_{s,t}e_{s,t}-\frac{J_t}{l_{f,t}}l_{m,t}}\\
	\nonumber
	\textrm{s.t.}  \quad &  \eqref{conl}, \eqref{cong}, \eqref{congr}, e_{s,t}\ge 0\\
	\label{xrg}
	&x_{i,\min}\le x_{i,t}\le \min\{a_{i,t},x_{i,\max}\}\\
	\label{xge}
	& \hspace{-1.5cm}\textstyle{x_{i,t}\ge \sum_{i=1}^Na_{i,t}-\sum_{j\neq i}^Nx_{j,t}-l_{m,t}+g_t-e_{s,t}}.
	\end{align}
	We solve \textbf{P3a)} by the partitioning method. Specifically, we first fix the variables $\big((x_{j,t})_{j\neq i}, l_{m,t}, g_t, e_{s,t}\big)$ and minimize \textbf{P3a)} over $x_{i,t}$. Since the objective function of \textbf{P3a)} is separable over all variables, an optimal solution of  $x_{i,t}$ can be derived by the following problem:
	\begin{align}\nonumber
	\min_{x_{i,t}} \quad &  VD_i(x_{i,t})+(s_{i,t}-\beta_i)x_{i,t} +   Vp_{b,t}x_{i,t}\\
	\nonumber
	\st \quad & \eqref{xrg}, \eqref{xge}.
	\end{align}
	
	Under the assumption that $s_{i,t}< \beta_i-V(p_{b,\max}+D'_{i,\max})= -x_{i,\min}+s_{i,\min}$, the objective function above is strictly decreasing with respect to $x_{i,t}$.
	Therefore, the optimal solution of $x_{i,t}$ is $\min\{a_{i,t},x_{i,\max}\}$.
	
	The demonstration of Lemma \ref{lem:sit}.2 is similar to that of Lemma \ref{lem:sit}.1.
	We first  transform \textbf{P3} to an equivalent problem \textbf{P3b)}  by eliminating the variables $e_{s,t}$ and $b_{i,t}, \forall i$, and the constant terms. To solve the problem, we first fix the variables $\big((x_{j,t})_{j\neq i}, l_{m,t}, g_t, e_{b,t}\big)$ and minimize \textbf{P3b)} over $x_{i,t}$. By some arrangement, an optimal solution of  $x_{i,t}$ can be derived by the following problem:
	\begin{align*}
	\min_{x_{i,t}} \quad &  VD_i(x_{i,t})+(s_{i,t}-\beta_i)x_{i,t} +   Vp_{s,t}x_{i,t}\\
	\st \quad & \eqref{xrg}\\
	&\textstyle{x_{i,t}\le \sum_{i=1}^Na_{i,t}-\sum_{j\neq i}^Nx_{j,t}-l_{m,t}+g_t+e_{b,t}}.
	\end{align*}
	When $s_{i,t}>\beta_i-V(p_{s,\min}+D'_{i,\min})$, the objective function above is strictly increasing with respect to $x_{i,t}$.
	Therefore, the optimal solution of $x_{i,t}$ is $x_{i,\min}$.
\end{IEEEproof}

Using Lemma \ref{lem:sit}, we can show that constraint \eqref{cons} holds by mathematical induction.
\begin{lemma}\label{lem:st}
	For the $i$-th storage unit, the energy state $s_{i,t}$ is  bounded within the interval $[s_{i,\min}, s_{i,\max}]$.
\end{lemma}
\begin{IEEEproof}
	The basis: For $t = 0$,  we have $s_{i,0}\in[s_{i,\min}, s_{i,\max}]$ for the initial setup.
	
	The inductive step: Assume that $s_{i,t}\in[s_{i,\min}, s_{i,\max}]$. Then we need to show that $s_{i,t+1}\in[s_{i,\min}, s_{i,\max}]$. In the following, we discuss  three cases of $s_{i,t}$.
	\begin{enumerate}[a)]
		\item $s_{i,t}\in[s_{i,\min},-x_{i,\min}+s_{i,\min})$.
		Using Lemma \ref{lem:sit}.1) and the iteration of $s_{i,t}$ in \eqref{st}, we have $s_{i,t+1} = s_{i,t}+\min\{a_{i,t},x_{i,\max}\}\ge s_{i,t}\ge s_{i,\min}$. Also, we have  $s_{i,t+1}\le s_{i,t}+x_{i,\max}< s_{i,\max}$ where the last inequality is derived based on the assumption of $s_{i,t}$ and $V_{\max}>0$.
		\item  $s_{i,t}\in[-x_{i,\min}+s_{i,\min},\beta_i-V(p_{s,\min}+D'_{i,\min})]$. Based on the iteration in \eqref{st}, we have $s_{i,t+1}\in[s_{i,t}+x_{i,\min},s_{i,t}+x_{i,\max}]$. By the definitions of $\beta_i$ and $V_{\max}$ we can derive that $s_{i,t+1}\in [s_{i,\min},s_{i,\max}]$.
		\item  $s_{i,t}\in(\beta_i-V(p_{s,\min}+D'_{i,\min}),s_{i,\max}]$. Using Lemma \ref{lem:sit}.2) and the iterations in \eqref{st}, we have $s_{i,t+1} = s_{i,t}+x_{i,\min}< s_{i,t}\le s_{i,\max}$. Also, we have $s_{i,t+1}>s_{i,\min}$ according to the assumption of $s_{i,t}$ and the definition of $\beta_i$.
	\end{enumerate}
\end{IEEEproof}

\section{Proof of Theorem \ref{the:iid}}\label{app:theiid}
1)
Note that \textbf{P2} fits the standard Lyapunov optimization format (see Section 4.3 in \cite{bkneely} for  details of the standard format). The idea of showing performance of Algorithm \ref{alg:rt} is to connect  Algorithm \ref{alg:rt} with the algorithm for  \textbf{P2} that  is designed under the Lyapunov optimization framework. Before showing  performance  of Algorithm \ref{alg:rt}, we give two lemmas, which will be used later.

In the following lemma, we show the existence of a special algorithm for \textbf{P2}. Denote $\tilde{w}$ as the optimal system cost of \textbf{P2}.
\begin{lemma}\label{lem:optst}
	For \textbf{P2}, there exists a stationary and randomized solution $\bu_t^s$ that only depends on the system states $\bq_t$, and at the same time satisfies the following conditions:
	\begin{align}
	\label{swt}
	&\E[w_t^s] \le \tilde{w}, \quad \forall t,\\
	\label{sxt}
	&\E[x_{i,t}^s] = 0, \quad \forall i, t,\\
	\label{slt}
	&\E\left[\frac{l_{b,t}+l_{f,t}-l_{m,t}^s}{l_{f,t}}\right] \le \alpha, \quad \forall t
	\end{align}
\end{lemma}
where  all expectations are taken over the randomness of the system state and the possible randomness of the decisions.
\begin{IEEEproof}
	The claims above can be derived from Theorem 4.5 in \cite{bkneely}. In particular, that theorem provides sufficient conditions for the existence of a stationary and randomized algorithm as described above. It can be checked that these sufficient conditions are all met in our problem. Therefore, the conclusion in Lemma \ref{lem:optst} holds.
\end{IEEEproof}

By minimizing the upper bound of the drift-plus-cost function (\ie, the right hand side of \eqref{dppup}), the real-time sub-problem for \textbf{P2} at time slot $t$ is given by
\begin{align}
\nonumber
\textbf{P3'}:\;\min_{\bu_t} \quad &\left[\sum_{i=1}^NVD_i(x_{i,t})+(s_{i,t}-\beta_i)x_{i,t}\right] + VC(g_t) \\
\nonumber
&+ Vp_{b,t}e_{b,t}-Vp_{s,t}e_{s,t}-\frac{J_t}{l_{f,t}}l_{m,t}\\
\nonumber
\textrm{s.t.}  \quad &
\eqref{conl}, \eqref{conx}, \eqref{conb}-\eqref{cong}, \eqref{cone}, \eqref{conmbl}.
\end{align}
Note that \textbf{P3'}  is the same as \textbf{P3} except without the ramping constraint \eqref{congr}.  Denote the optimal objective values of \textbf{P3'} and \textbf{P3} as $\tilde{f}_t$ and $f_t^*$, respectively, and denote  an optimal solution of \textbf{P3'} and \textbf{P3} as $\tilde{\bu}_t$ and $\bu_t^*$, respectively.  In the following lemma, we characterize $f_t^*$ in terms of $\tilde{f}_t$.

\begin{lemma}\label{lem:gr}
	At each time slot,  $f_t^*$ is bounded as
	$\tilde{f}_t\le f_{t}^*\le \tilde{f}_t +\epsilon$, where \[\epsilon\define  V(1-r)g_{\max}\max\{p_{b,\max},C'_{\max}\}.\]
\end{lemma}
\begin{IEEEproof}
	First, since \textbf{P3} has more  restricted constraints than \textbf{P3'}, there is $f_t^*\ge \tilde{f}_t$.
	
	Next, we are to upper bound $f_t^*-\tilde{f}_t$.
	Comparing the solution $g_t^*$ of \textbf{P3} with the solution $\tilde{g}_t$ of \textbf{P3'}  there are  three possibilities:
	\begin{enumerate}
		\item $g_t^*=\tilde{g}_t$,
		\item $g_t^* < \tilde{g}_t$ (less output due to  constraint \eqref{congr}), and
		\item $g_t^* >\tilde{g}_t$ (more output due to  constraint \eqref{congr}).
	\end{enumerate}
	For Case 1), it is easy to show that $f_t^*=\tilde{f}_t$. Thus, we  focus on the latter two cases.
	
	Denote  a feasible solution of \textbf{P3} as $\hat{\bu}_t$ and its corresponding  objective value  as $\hat{f}_t$. Since characterizing the gap $f_t^*-\tilde{f}_t$ directly is challenging, we instead consider the gap  $\hat{f}_t-\tilde{f}_t$.
	
	For Case 2), when $g_t^*<\tilde{g}_t$, the effective constraint of $g_t$ in \textbf{P3} should be $\max\{g_{t-1}-rg_{\max},0\}\le g_t\le g_{t-1}+ rg_{\max}$. Set a feasible solution of  \textbf{P3} as $\hat{\bu}_t = [\tilde{\bb}_{t}, \tilde{\bx}_{t},\tilde{l}_{m,t}, g_{t-1}+rg_{\max},\tilde{e}_{b,t}+\tilde{g}_t-g_{t-1}-rg_{\max},\tilde{e}_{s,t}].$ That is, $\hat{\bu}_t$ is the same as $\tilde{\bu}_t$ except the solutions of $g_t$ and $e_{b,t}$. Intuitively, we can interpret  $\hat{\bu}_t$ as that, due to the ramping constraint, the CG is forced  to generate less energy, and  the aggregator  chooses to  buy more  from the external energy markets to balance power. The gap $\hat{f}_t-\tilde{f}_t$ is given by
	\begin{align}
	\nonumber
	&\hat{f}_t-\tilde{f}_t \\
	\nonumber
	&= V\big[C(g_{t-1}+rg_{\max})-C(\tilde{g}_t)+p_{b,t}(\tilde{g}_t-g_{t-1}-rg_{\max}) \big]\\
	\label{gt21}
	&\le Vp_{b,t}(\tilde{g}_t-g_{t-1}-rg_{\max})\\
	\label{gt211}
	&\le V(1-r)g_{\max}p_{b,\max}
	\end{align}
	where the inequality in \eqref{gt21} holds since $\tilde{g}_t>g_{t-1}+rg_{\max}$ and the function $C(\cdot)$ is non-decreasing.
	From \eqref{gt211}, the gap $f_t^*-\tilde{f}_t$ is upper bounded by
	\begin{align}\label{gt22}
	f_t^*-\tilde{f}_t\le \hat{f}_t-\tilde{f}_t\le V(1-r)g_{\max}p_{b,\max}.
	\end{align}
	
	The proof for  Case 3) is similar as that for Case 2). In particular,  when $g_t^*>\tilde{g}_t$, the effective constraint of $g_t$ in \textbf{P3} should be $g_{t-1}-rg_{\max} \le g_t\le \min\{g_{\max}, g_{t-1}+rg_{\max}\}$. Set a feasible solution of  \textbf{P3} as $\hat{\bu}_t = [\tilde{\bb}_{t}, \tilde{\bx}_{t},\tilde{l}_{m,t}, g_{t-1}-rg_{\max},\tilde{e}_{b,t},\tilde{e}_{s,t}-\tilde{g}_t+g_{t-1}-rg_{\max}].$ That is, $\hat{\bu}_t$ is the same as $\tilde{\bu}_t$ except the solutions of $g_t$ and $e_{s,t}$. Intuitively, we can interpret $\hat{\bu}_t$ as that, due to the ramping  constraint, the CG is forced to generate more energy, and  the aggregator chooses to  sell more  to the external energy markets to balance power. The gap $\hat{f}_t-\tilde{f}_t$ is given by
	\begin{align}
	\nonumber
	&\hat{f}_t-\tilde{f}_t \\
	\nonumber
	&=V\big[C(g_{t-1}-rg_{\max})-C(\tilde{g}_t)+p_{s,t}(\tilde{g}_t-g_{t-1}+rg_{\max}) \big]\\
	\label{gt31}
	&\le V\big[C(g_{t-1}-rg_{\max})-C(\tilde{g}_t)\big]\\
	\label{gt32}
	&\le V(g_{t-1}-rg_{\max}-\tilde{g}_t)C'_{\max}\\
	\label{gt322}
	&\le V(1-r)g_{\max}C'_{\max}
	\end{align}
	where the inequality in \eqref{gt31} holds since $\tilde{g}_t<g_{t-1}-rg_{\max}$, and the inequality \eqref{gt32} is derived by the mean value theorem.
	From \eqref{gt322}, we have
	\begin{align}\label{gt33}
	f_t^*-\tilde{f}_t\le \hat{f}_t-\tilde{f}_t\le V(1-r)g_{\max}C'_{\max}.
	\end{align}
	Combining  \eqref{gt22} and \eqref{gt33} yields  $f_t^*\le \tilde{f}_t + V(1-r)g_{\max}\max\{p_{b,\max},C'_{\max}\}$, which completes the proof.
\end{IEEEproof}

Using Lemmas \ref{lem:dpp}, \ref{lem:optst}, and  \ref{lem:gr}, the drift-plus-cost function under Algorithm \ref{alg:rt} can be upper bounded below:
\begin{align}
\nonumber
&\Delta(\mb{\Theta}_t)+V\E[{w}_t^*|\mb{\Theta}_t]\\
\nonumber
&\le B+\epsilon+J_t\E\left[\frac{l_{b,t}+l_{f,t}-\tilde{l}_{m,t}}{l_{f,t}}-\alpha\Big|\mb{\Theta}_t\right]\\
\label{lyth0}
&\hspace{1cm}+ \sum_{i=1}^N(s_{i,t}-\beta_i)\E\big[\tilde{x}_{i,t}|\mb{\Theta}_t\big]+V \E[\tilde{w}_t|\mb{\Theta}_t]\\
\nonumber
&\le B+\epsilon+J_t\E\left[\frac{l_{b,t}+l_{f,t}-l_{m,t}^s}{l_{f,t}}-\alpha\Big|\mb{\Theta}_t\right]\\
\label{lyth1}
&\hspace{1cm}+ \sum_{i=1}^N(s_{i,t}-\beta_i)\E\big[x_{i,t}^s|\mb{\Theta}_t\big]+V \E[w_t^s|\mb{\Theta}_t]\\
\label{lyth2}
&\le B+\epsilon+V\tilde{w} \\
\label{lyth3}
&\le B+\epsilon+Vw^{\opt}
\end{align}
where  \eqref{lyth0} is derived by Lemmas \ref{lem:dpp} and \ref{lem:gr},    \eqref{lyth1} holds since \textbf{P3'} minimizes the right hand side of  \eqref{lyth0}, \eqref{lyth2} is derived based on \eqref{swt}\eqref{sxt}\eqref{slt} in Lemma \ref{lem:optst} and the fact that $\bu_t^s$ is independent of $\mb{\Theta}_t$, and  \eqref{lyth3} holds since \textbf{P2} is a relaxed problem of  \textbf{P1}.

Taking  expectations over $\mb{\Theta}_t$ on both sides of \eqref{lyth3} and summing over $t\in \{0,\cdots, T-1\}$ yields
\begin{align}\label{lydf1}
\E[L(\mb{\Theta}_T)]-\E[L(\mb{\Theta}_0)]+V\sum_{t=0}^{T-1}\E[{w}_t^*]\le (B+\epsilon+Vw^{\opt})T.
\end{align}
Since $L(\mb{\Theta}_T)$ is non-negative, after some arrangement, from \eqref{lydf1} there is
\begin{align}\label{lydf2}
\frac{1}{T}\sum_{t=0}^{T-1}\E[{w}_t^*]\le \frac{B+\epsilon+Vw^{\opt}}{V} + \frac{\E[L(\mb{\Theta}_0)]}{TV}.
\end{align}
Taking $\limsup$ on both sides of \eqref{lydf2} and rearranging the terms gives
$w^*-w^{\opt} \le  B/V+(1-r)g_{\max}\max\{p_{b,\max},C'_{\max}\}.$ To emphasize the dependence of performance on $r$ and $V$, we express $w^*$ as $w^*(r, V)$. Similarly, we express $w^{\opt}$ as $w^{\opt}(r)$.

2)   The lower bound on $w^{\opt}(r)$ can be derived by setting $r = 1$ in  Theorem \ref{the:iid}.1 and recognizing that $w^{\opt}(1)\le w^{\opt}(r)$.

\section{Proof of Proposition \ref{pro:st}}\label{app:prost}
Proposition \ref{pro:st} can be shown by mathematical induction. The proof resembles that of Lemma \ref{lem:st} where the energy capacity $s_{i,\max}$ is replaced by  $s_{i,\textrm{up}}$. We omit the proof for brevity.

\section{Proof of Proposition \ref{pro:prop}}\label{app:proprop}
1) We prove the conclusion by mathematical induction.

The basis: For $t = 0$,  we have $J_t = 0$, which is obviously  upper bounded.

The inductive step:  Assume that $J_t\le Vp_{b,\max}l_{f,\max}+1$. Then we need to show that $J_{t+1}\le Vp_{b,\max}l_{f,\max}+1$. Consider the following two cases of $J_t$.
\begin{enumerate}[a)]
	\item $J_t\le Vp_{b,\max}l_{f,\max}$. Based on the update of $J_t$ in \eqref{qj}, we have $J_{t+1}\le \max\{J_t-\alpha,0\} + 1\le J_t+1\le Vp_{b,\max}l_{f,\max}+1$.
	\item $J_t\in(Vp_{b,\max}l_{f,\max}, Vp_{b,\max}l_{f,\max}+1]$. For this case, we will show  that the unique solution of $l_{m,t}$ to \textbf{P3} is $l_{b,t}+l_{f,t}$. Hence, $J_{t+1} = \max\{J_t-\alpha,0\}\le J_t\le Vp_{b,\max}l_{f,\max}+1$.
	
	To this end, we consider the equivalent problem \textbf{P3a)}. First fix the  variables $\big(\bx_t,  g_t, e_{s,t}\big)$ and minimize \textbf{P3a)} over $l_{m,t}$. After some arrangement, an optimal solution of  $l_{m,t}$ can be derived by the following problem:
	\begin{align*}
	\min_{l_{m,t}} \quad & \big(Vp_{b,t}-\frac{J_t}{l_{f,t}}\big)l_{m,t}\\
	\st \quad & l_{b,t}\le l_{m,t}\le l_{b,t}+l_{f,t},\\
	& l_{m,t}\ge \sum_{i=1}^N(a_{i,t}-x_{i,t})+g_t-e_{s,t}.
	\end{align*}
	When $J_t>Vp_{b,\max}l_{f,\max}$, the objective function above is strictly decreasing. Therefore, the optimal solution of $l_{m,t}$ is $l_{b,t}+l_{f,t}$.
	
\end{enumerate}

2) We prove the conclusion by contradiction. Suppose that under our algorithm the optimal solutions of $e_{b,t}$ and $e_{s,t}$ satisfy $e_{b,t}^*>e_{s,t}^*>0$. Then, we can show that there is another feasible solution $\hat{\bu}_t = \left[\bb_t^*,  \bx_t^*, l_{m,t}^*, g_t^*, e_{b,t}^*-e_{s,t}^*, 0\right]$ achieving a strictly smaller objective value, hence contradicting the fact that $\bu_t^*$ is optimal. The proofs of the other two possible cases, \ie, $e_{b,t}^*=e_{s,t}^*>0$ and $e_{s,t}^*>e_{b,t}^*>0$, are similar, and are omitted  for brevity.

\section{Simplification of  \eqref{yiup}-\eqref{diup} } \label{app:deradmm}
Define  $\overline{y}^k\define\frac{1}{N+4} \sum_{i=1}^{N+4}y_i^k$ and $\overline{d}^k\define\frac{1}{N+4} \sum_{i=1}^{N+4}d_i^k$ as the averages of $y_i^k$ and $d_i^k$ over $i$ at the $k$-th iteration, respectively.
By solving the minimization problem in \eqref{zup},  we can get a closed-form solution of  $z_i^{k+1}$ below:
\begin{align}\label{ziup}
z_i^{k+1} = \frac{d_i^k}{\rho}+y_i^{k+1}-\frac{\overline{d}^k}{\rho}-\overline{y}^{k+1}+\frac{\sum_{i=1}^Na_i}{N+4}.
\end{align}

Substituting the  right hand side of \eqref{ziup} for $z_i^{k+1}$ in the $\bd$-update  \eqref{diup} yields $d_i^{k+1} = \overline{d}^k+\rho(\overline{y}^{k+1}-\frac{\sum_{i=1}^Na_i}{N+4})$, which indicates that the dual variables $d_i^{k+1}$ are identical for all $i$ at each iteration. Therefore, we can safely drop the subscript $i$ in $d_i^{k+1}$ and obtain the $d$-update in \eqref{dup}. Meanwhile, substituting the right hand side of \eqref{ziup} for $z_i^k$ in the $\by$-update  \eqref{yiup} and using the fact that $d_{i}^{k-1}$ are identical for all $i$ yields \eqref{yupf}. Since the vector $\bz$ is not employed in either $\by$-update or $d$-update, it can be eliminated.

\begin{IEEEbiography}[{\includegraphics[width=1in,height=1.25in,clip,keepaspectratio]{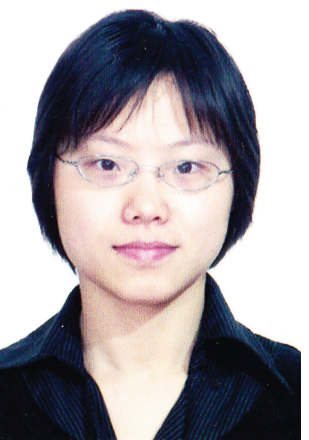}}]
	{Sun Sun} (S'11)
	received the B.S. degree in Electrical Engineering and Automation  from  Tongji University, Shanghai, China, in 2005. From 2006 to 2008, she was a software engineer in the Department of GSM Base Transceiver Station of Huawei Technologies Co. Ltd.. She received the M.Sc. degree in Electrical and Computer Engineering from University of Alberta, Edmonton, Canada, in 2011.
	Now, she is pursuing her Ph.D. degree in the Department of Electrical and Computer Engineering of University of Toronto, Toronto, Canada.
	Her current research interest lies in the areas of stochastic optimization, distributed control, learning, and economics, with the application of renewable generation, energy storage,
	demand response, and  power system operations.
\end{IEEEbiography}

\begin{IEEEbiography}[{\includegraphics[width=1in,height=1.25in,clip,keepaspectratio]{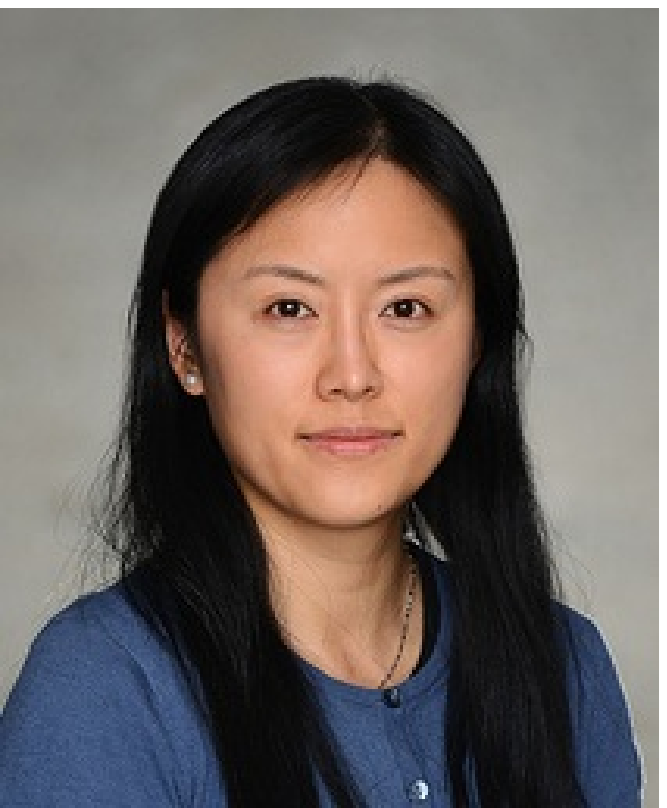}}]
{Min Dong} (S'00-M'05-SM'09) received the B.Eng. degree from Tsinghua University, Beijing, China, in 1998, and the Ph.D. degree in electrical and computer engineering with minor in applied mathematics from Cornell University, Ithaca, NY, in 2004. From 2004 to 2008, she was with Corporate Research and Development, Qualcomm Inc., San Diego, CA. In 2008, she joined the Department of Electrical, Computer and Software Engineering at University of Ontario Institute of Technology, Ontario, Canada, where she is currently an Associate Professor. She also holds a status-only Associate Professor appointment with the Department of Electrical and Computer Engineering, University of Toronto since 2009. Her research interests are in the areas of statistical signal processing for communication networks, cooperative communications and networking techniques, and stochastic network optimization in dynamic networks and systems. She served as an Associate Editor for the IEEE TRANSACTIONS ON SIGNAL PROCESSING (2010–2014), and as an Associate Editor for the IEEE SIGNAL PROCESSING LETTERS (2009–2013). She was a technical lead co-chair of the Communications and Networks to Enable the Smart Grid Symposium at the IEEE International Conference on Smart Grid Communications (SmartGridComm) in 2014. She has been an elected member of the IEEE Signal Processing Society Signal Processing for Communications and Networking (SP-COM) technical committee since 2013. She was the recipient of the Early Researcher Award from Ontario Ministry of Research and Innovation in 2012, the Best Paper Award at IEEE ICCC in 2012, and the 2004 IEEE Signal Processing Society Best Paper Award.
\end{IEEEbiography}

\begin{IEEEbiography}[{\includegraphics[width=1in,height=1.25in,clip,keepaspectratio]{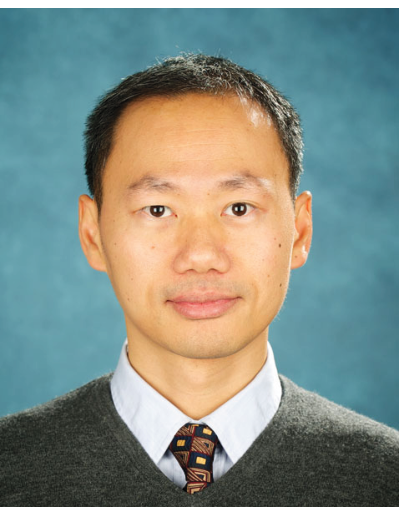}}]
	{Ben Liang}(S'94-M'01-SM'06)
	received honors-simultaneous B.Sc. (valedictorian) and M.Sc. degrees in Electrical Engineering from Polytechnic University in Brooklyn, New York, in 1997 and the Ph.D. degree in Electrical Engineering with Computer Science minor from Cornell University in Ithaca, New York, in 2001. In the 2001 - 2002 academic year, he was a visiting lecturer and post-doctoral research associate at Cornell University. He joined the Department of Electrical and Computer Engineering at the University of Toronto in 2002, where he is now a Professor. His current research interests are in mobile communications and networked systems. He has served as an editor for the IEEE Transactions on Wireless Communications and an associate editor for the Wiley Security and Communication Networks journal, in addition to regularly serving on the organizational or technical committee of a number of conferences. He is a senior member of IEEE and a member of ACM and Tau Beta Pi.
\end{IEEEbiography}

\end{document}